\documentclass[twocolumn]{aastex631}
\received{2024 February 29}
\accepted{2024 April 18}

\usepackage{longtable}
\usepackage{graphics}
\usepackage{epsf}
\usepackage{graphicx}	
\usepackage{amsmath}	
\usepackage{amssymb}	
\usepackage{hyperref}

\shorttitle{IC 3599}
\shortauthors{D. Grupe et al.}

\begin{document}

\title{
The calm before the (next) storm: no third outburst in 2019--2020, and ongoing monitoring of the transient AGN IC 3599
}

 \author[0000-0002-9961-3661]{Dirk Grupe}
 \affiliation{Northern Kentucky University, 
Department of Physics, Geology, and Engineering Technology,
Nunn Drive, 
Highland Heights, KY 41099, USA }

\author[0000-0003-4183-4215]{S. Komossa}
\affiliation{Max-Planck-Institut f{\"u}r Radioastronomie, 
Auf dem H{\"u}gel 69, 53121 Bonn, Germany}

 \author{Salem Wolsing}
 \affiliation{Northern Kentucky University, 
Department of Physics, Geology, and Engineering Technology,
Nunn Drive, 
Highland Heights, KY 41099, USA }

\newcommand{\swift}{{\it Swift}}
\newcommand{\suzaku}{{\it Suzaku}}
\newcommand{\xmm}{{\it XMM-Newton}}
\newcommand{\chandra}{{\it Chandra}}
\newcommand{\ax}{$\alpha_{\rm X}$}
\newcommand{\rb}[1]{\raisebox{1.5ex}[-1.5ex]{#1}}
\newcommand{\msun}{$M_{\odot}$}
\newcommand{\dM}{\dot M}
\newcommand{\dMM}{$\dot{M}/M$}
\newcommand{\dMedd}{\dot M_{\rm Edd}}
\newcommand{\plm}{$\pm$}
\newcommand{\nh}{$N_{\rm H}$}
\newcommand{\auv}{$\alpha_{\rm UV}$}
\newcommand{\aox}{$\alpha_{\rm ox}$}
\newcommand{\wpvs}{{WPVS~007}}
\newcommand{\lledd}{$L/L_{\rm Edd}$}

\begin{abstract}
We report on follow-up observations of the Seyfert 1.9 galaxy IC 3599 with the NASA Neil Gehrels \swift\ mission. 
The detection of a second X-ray outburst in 2010 by \swift\ after the first discovery of a bright X-ray outburst in 1990 by ROSAT led to 
the suggestion of two very different explanations: The first one
assumed that IC 3599 exhibits outbursts due to repeated partial tidal stripping of a star,  
predicting another outburst of IC 3599 in 2019/2020. The second, alternative scenario assumed that the event observed in X-rays is due to an accretion disk instability which would suggest a much longer period between the large outbursts. Our continued monitoring campaign by \swift\ 
allowed us to test  the first scenario which predicted a repetition of high amplitude flaring activity in 2019/2020. 
We do not find any evidence of dramatic flaring activity with factors of 100
since the last X-ray outburst seen in 2010. These
observations support the accretion disk scenario. 
Further, while IC 3599 remains in low emission states, the long-term X-ray light curve of IC 3599 reveals ongoing strong variability of a factor of a few. The most remarkable event is a mini flare of a factor of 10 in X-rays in December 2022. After that flare, the otherwise supersoft X-ray spectrum shows an exceptional hardening, reminiscent of a temporary corona formation.
\end{abstract}


\section{Introduction}

Since the proposed scenario of the tidal disruption of a stars by a supermassive black holes \citep[e.g.,][]{rees1988}, there have been many observations confirming this scenario. 
In the 1990s the X-ray mission ROSAT \citep{truemper1982} had been instrumental for the discovery of the first stellar tidal disruption events (TDEs), in the form of giant-amplitude X-ray outbursts from quiescent host galaxies
(e.g., NGC 5905 \citep{bade1996, komossa1999a},  RX J624.9+7554 \citep{grupe1999},  and RX J1242.6--1119 \citep{komossa1999b}). More recently, TDEs have been detected by all major X-ray observatories, including the discovery of jetted events with the Neil Gehrels Swift observatory, first in Swift\,J164449.3+573451 \citep[e.g.,][]{burrows2011, Bloom2011, Zauderer2011}.
 The discovery of TDEs from quiescent galaxies, however, is not limited to the X-ray sky. GALEX led to the first detection of events in the UV-optical band \citep[][]{gezari2006}, 
 and SDSS enabled the first identification of events with transient  optical emission lines such as HeII, broad Balmer lines, and coronal lines \citep{Komossa2008}. 
 TDE candidates continue to be identified
 at all wavebands \citep[see][for recent reviews]{Komossa2015, vanvelzen2021, webb2023, komossa2023}.  
 
 Although the TDE model is required to explain sudden, luminous X-ray outbursts in {\em{non-active}} galaxies, this picture is far from being clear in AGN. In the case of an AGN, the power is coming from a long-lived accretion disk surrounding the central supermassive black hole. This power is not constant and varies, and in extreme cases a significant change in the disk properties 
 can cause a dramatic increase in the AGN luminosity. An increasing number of optical changing-look AGN which do change their luminosity output along with their optical broad emission lines, have been identified in recent years \citep[see][for a review]{komossa2023} even though the first cases were known already in the 1970s--80s \citep[e.g.,][]{Penston1984, alloin1985}. 
 Different theoretical scenarios related to accretion disk instabilities or other disk processes have been explored in recent years
 \citep[e.g.,][]{Nicastro2000, grupe2015, ross2018, NodaDone2018, DexterBegelman2019, Sniegowska2020, PanXin2021, Kaaz2023, Cao2023}.
 There will be a major discovery space for both, TDEs and changing-look AGN once the Vera Rubin Telescope will go online{\footnote{https://rubinobservatory.org/}}.

One of the earliest examples of an AGN that might have exhibited either a TDE or an accretion disk instability was the  Seyfert 1.9 galaxy IC 3599 
(Zw 159.034; 1RXS J123741.2$+264227$;
$\alpha_{2000}$ = $12^{\rm h} 37^{\rm m} 41.^{\rm s}2$, 
$\delta_{2000}$ = $+26^{\circ} 42' 27^{''}$, z=0.0215). IC 3599 was discovered as a very X-ray bright AGN during the ROSAT
All-Sky Survey \citep[RASS,][]{voges1999} in 1990. Subsequent pointed ROSAT observations during the following years revealed that it had faded by a factor of initially 60 and even more than 100 in later observations. 
Both, \citet{brandt1995} and \citet{grupe1995a} speculated at the time that this was caused by either a TDE or high-amplitude AGN variability. 

What made both scenarios plausible was the discovery of a strong evolution in the optical emission line spectrum. The spectrum taken about half a year after the RASS observation shown in \citet{brandt1995} showed very strong permitted lines from Hydrogen and Helium along with transitions from other elements. 
 These strong emission lines led \citet{brandt1995} to classify IC 3599 as a Narrow-Line Seyfert 1 galaxy (NLS1), albeit without FeII emission. \citet{grupe1995a} showed spectra obtained years after the RASS which displayed the spectrum of a Seyfert 2 galaxy but with fading coronal iron lines. 
 The fading coronal lines were confirmed by new optical spectroscopy of  \citet{komossa1999a} who still detected a broad component in H$\alpha$, making IC 3599 a Seyfert 1.9 galaxy. These authors re-emphasized that TDEs are best identified in quiescent host galaxies, because in AGN, accretion-disk-related activity is the most plausible explanation for variability \citet[see also][]{rees1990}.
 Note, no FeII lines, the signature lines in Narrow Line Seyfert 1 galaxies, were present at any time in the optical data of IC 3599. 
Doubt remained given that this is an AGN, if a TDE really was an explanation of the X-ray outburst.  Instead, the spectacular changes in its optical emission-line spectrum combined with the dramatic X-ray variability make IC 3599 one of the most extreme examples of a changing-look AGN \citep{komossa2023}. 
 
After ROSAT,  IC 3599 was again observed in X-rays by Chandra in 2002  \citep{vaughan2004} and was still found in an X-ray low state. In two observations by the Neil Gehrels 
\swift\ mission \citep[][\swift\ hereafter]{gehrels2004}  in 2010, it was seen again in an X-ray outburst state with its X-ray flux at a similar level as during the RASS \citep{grupe2015}. Based on the RASS observation from 1991, the Chandra observation from 2002, and the Swift observation from 2010, \citet{campana2015} speculated that the 2010 outburst was due to the 
 tidal stripping of a star, repeating an initial 1991 tidal stripping event  of a star orbiting the black hole, with a low black hole mass of $3\times 10^5 M_\odot$
and a 9.5-year orbital period. This kind of repeated flaring has been suggested e.g. by \citet{payne2021} to explain the quasi-periodic eruptions in ESO 253-G003/AT2014ko. 
\citet{campana2015} predicted that another outburst in IC 3599 would be visible in 2019/2020. Although this is an intriguing model, \citet{grupe2015} argued that the 1990 and the 2010 outbursts were due to accretion disk instabilities. Major arguments were the optical light curve from the Catalina observatory which showed a relatively gradual increase in the optical flux,  as well as a low black hole mass not being supported by any observational parameter (instead, it was measured to be of the order of $2-20\times 10^6 M_\odot$). 

 To test these different models, we continued to monitor IC 3599 with Swift typically once per month for a nine-month period each year. \swift\ is unable to observe IC 3599 for a three-month period between August and November due to the sun-constraint (a 45$^{\circ}$ sun avoidance angle). Here we report on the entire \swift\ monitoring campaign from 2010 to June 2023.  During the entire monitoring period no further outbursts have been observed in IC 3599, supporting the accretion disk instability scenario suggested by \citet{grupe2015}. Over the last few years there has been, however, a general trend of IC 3599 slowly becoming brighter in X-rays, and it has recently been observed in an intermediate state.

This paper is organized as follows: in Section 2 we will describe the \swift\ observations, present the results in Section 3, and then discuss them in Section 4. 
Throughout the paper spectral indices are denoted as 
$F_{\nu} \propto \nu^{-\alpha}$. Luminosities are calculated assuming a $\Lambda$CDM
cosmology with $\Omega_{\rm M}$=0.286, $\Omega_{\Lambda}$=0.714 and a Hubble
constant of $H_0$=70 km s$^{-1}$ Mpc$^{-1}$, resulting in a luminosity
 distance of $D$=91.4 Mpc
using the cosmology calculator  by \citet{Wright2006}.
 All errors are 1$\sigma$ unless stated otherwise.

\section{\swift\ Observations and Data Reduction}

\swift\ has observed IC 3599 since 2010. When it was discovered in 2013 that it had exhibited an X-ray outburst in 2010 \citep{grupe2015} we started observing it once a month. These observations were supported by another set of monthly observations (PI Campana) in 2019 which essentially made it a two-week cadence. All \swift\ observations are listed 
in Table\,\ref{swift_obs}. 
The full machine-readable table is available on Zenodo: \dataset[10.5281/zenodo.10899673]{\doi{10.5281/zenodo.10899673}}.

The \swift\ X-ray telescope \citep[XRT;][]{burrows2005}
was operating in photon counting mode \citep[pc mode,][]{hill2004} and the
data were reduced by the task {\it xrtpipeline}
which is part the HEASOFT package 6.30.1. Source counts were extracted in a
circle with a radius of 47.1$^{''}$  (except the 2010 outburst observation for which we used 70.7$^{''}$). 
The
background counts were extracted in a nearby 
circular region with a radius of 247.5$^{''}$. 
For all spectra we used the most recent response file {\it swxpc0to12s6\_20130101v014.rmf}.
The X-ray spectra were analyzed using {\it XSPEC} version 12.12.1 \citep{arnaud1996}.  If a sufficient number of counts ($>$30) was available, we used 
 W-statistics \citep{cash1979} for single observations. Although often the number of counts was too low to allow for a spectral analysis, we still determined a hardness ratio,  defined as $HR = \frac{hard - soft}{hard+soft}$ with the soft and hard bands in the 0.3-1.0 and 1.0-10 keV range, respectively by applying the 
  program by \citet{park2006}. The count rates in the XRT were determined by using the \swift\, XRT product page at the \swift\, Data Center at the University of Leicester\footnote{\url{https://www.swift.ac.uk/user_objects/}} \citep{evans2007}, as well as the Living Swift XRT Point Source Catalogue \footnote{\url{https://www.swift.ac.uk/LSXPS/}} \citep{evans2023}. 
  
  To perform a spectral analysis of different X-ray flux states, we merged the data of various time periodd into single data sets within XSELECT. These periods and spectral analysis results are listed in the appendix in Table\,\ref{xrt_merge}. We then constructed the auxiliary response files (arf) by using the FTOOLS command {\it addarf}, using the weighting factor of each individual  arf by dividing the individual exposure time of each single observation by the total exposure time of the merged spectrum.

The UV-optical telescope \citep[UVOT;][]{roming2005}
data of each segment were co-added in each filter with the UVOT
task {\it uvotimsum}. In general, all observations were performed in all 6 UVOT filters, except if the observation was interrupted by Gamma-Ray Burst detections or higher priority Target of Opportunity observations. 
Source counts in all 6 UVOT filters
  were selected in a circle with a radius of 5$^{''}$ and background counts in
  a nearby source free region with a radius of 20$^{''}$.
  UVOT magnitudes and fluxes were measured with the task {\it  
uvotsource} based on the most recent UVOT calibration as described in  
\citet{poole2008} and  \citet{breeveld2010}.
The UVOT data were corrected for Galactic reddening
\citep[$E_{\rm B-V}=0.015$;][]{schlegel1998}. The correction factor in each  
filter was
calculated with equation (2) in \citet{roming2009}
who used the standard reddening correction curves by \citet{cardelli1989}.

\section{Results}

\subsection{X-ray and UV/Optical Light Curves}

\subsubsection{Light Curves}

The fluxes in X-rays and in the UV/optical obtained by \swift\ 
are listed in Table\,\ref{swift_res}.  This table also contains the observations published already in \citet{grupe2015}
which were re-analyzed due to updates in the calibration files. 
The full machine-readable table is available on Zenodo: \dataset[10.5281/zenodo.10899673]{\doi{10.5281/zenodo.10899673}}. Note that many 0.3-10 keV fluxes were estimated from the count rates derived from the Leicester \swift-XRT Product tool (see footnote 1) and then converted to fluxes based on the closest X-ray data that allowed a spectral analysis. These estimated fluxes are marked in Table\,\ref{swift_res}.

\begin{figure}
\includegraphics[trim=10 150 15 300,clip,width=9cm]{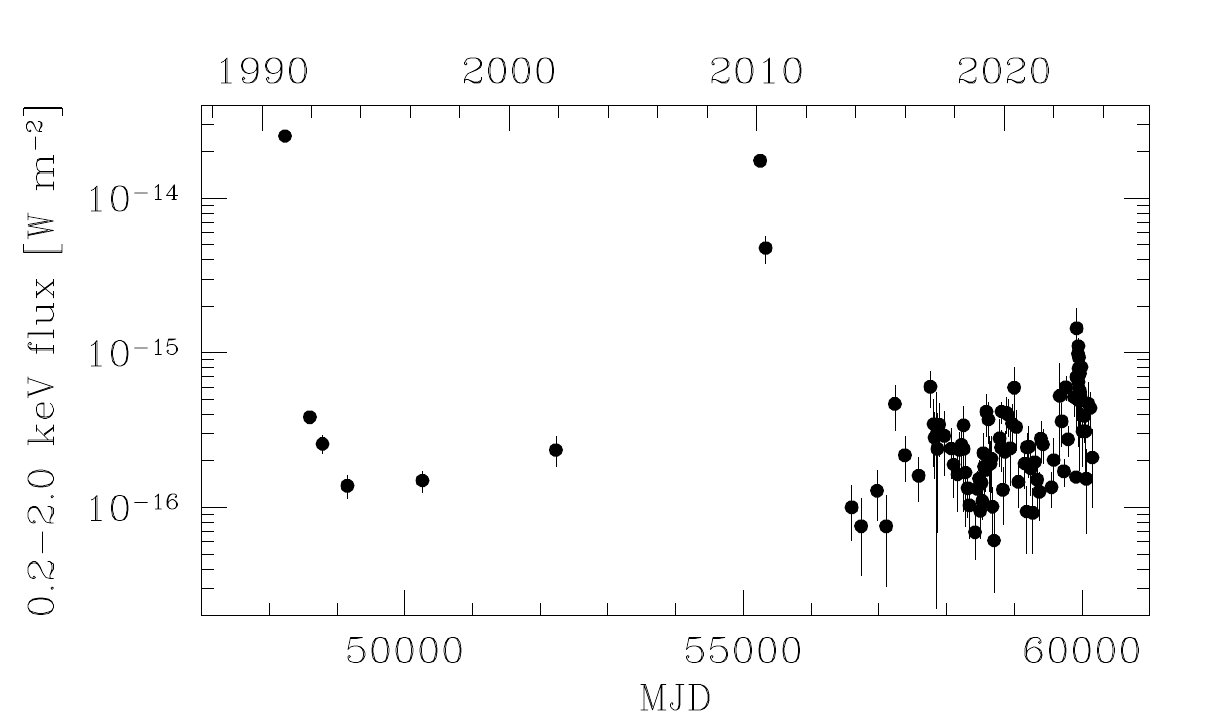}
    \caption{Long-term 0.2-2.0 keV X-ray light curve of IC 3599. The data during the 1990s are from ROSAT, the 2002 observation from Chandra, and all others after 2010 from Swift. Data are available at \dataset[10.5281/zenodo.10899673]{\doi{10.5281/zenodo.10899673}}  \label{xray_lc}}
\end{figure}

Figure\,\ref{xray_lc} displays the long-term X-ray light curve of IC 3599. The data for this light curve are available on Zenodo: \dataset[10.5281/zenodo.10899673]{\doi{10.5281/zenodo.10899673}}.
The light curves clearly show the two giant outbursts seen by ROSAT and \swift\ about 20 years apart. The long-term light curve also displays a mini flare which shows an increase in X-ray flux by a factor of about 10 in December 2022.

\begin{figure*}
\includegraphics[trim=10 150 15 15,clip,width=\textwidth]{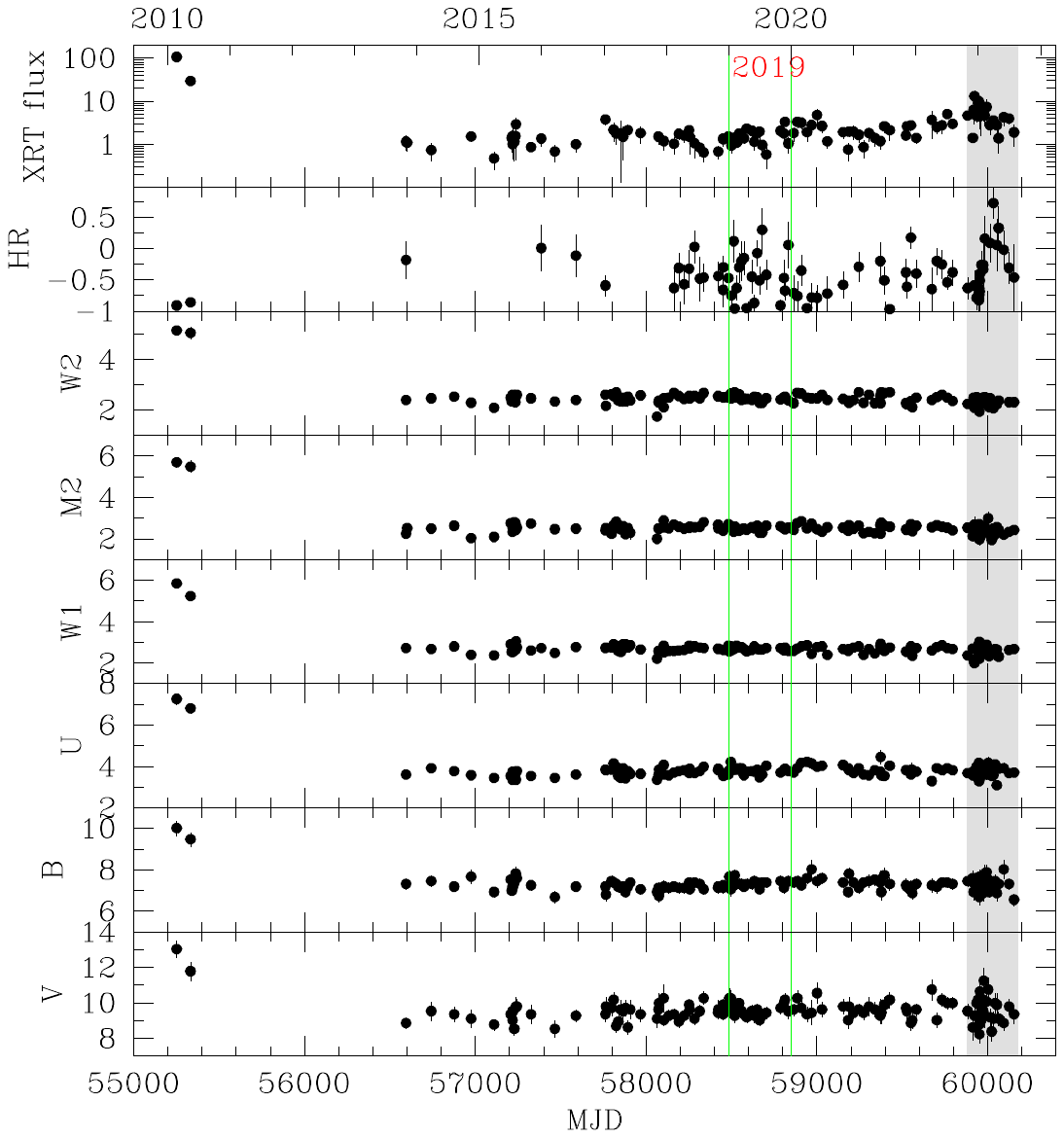}
    \caption{Observed Swift X-ray to optical light curve of IC 3599 from 2010 to August 2023.  The fluxes in the X-ray and UVOT bands are given in units of $10^{-16}$~ W m$^{-2}$. The grey-shaded area displays the range of the 2023 Swift light curve displayed in Figure \ref{swift_lc_2023}. 
    \label{swift_lc}}
\end{figure*}

Figure\ \ref{swift_lc} displays the full \swift\ light curve between 2010 and August 2023. 
Note that due to the low background in the \swift\ XRT even a low number of counts results in a 3$\sigma$ detection. However, the number of counts is often too low to allow for a reliable determination of the hardness ratio. 
It displays the outburst in 2010 and the remarkable flat light curve since \swift\ resumed the observations with our monitoring program in 2013. The green vertical lines mark the year 2019 when \citet{campana2015} predicted that the hypothetic  star would return and undergo another partial disruption event while going through periastron
and thus form a temporary accretion disk around the black hole again, causing another outburst. Clearly, this did not happen. There are no signs of unusual activity in X-rays or  UV, yet. There is, however, a general trend since about 2021 that IC 3599 is slowly becoming brighter in X-rays (see below in Section\,\ref{merged}).  The hardness-ratio light curve suggests strong spectral variability in X-rays as well.

\begin{figure*}
\includegraphics[trim=10 150 15 15,clip,width=\textwidth]{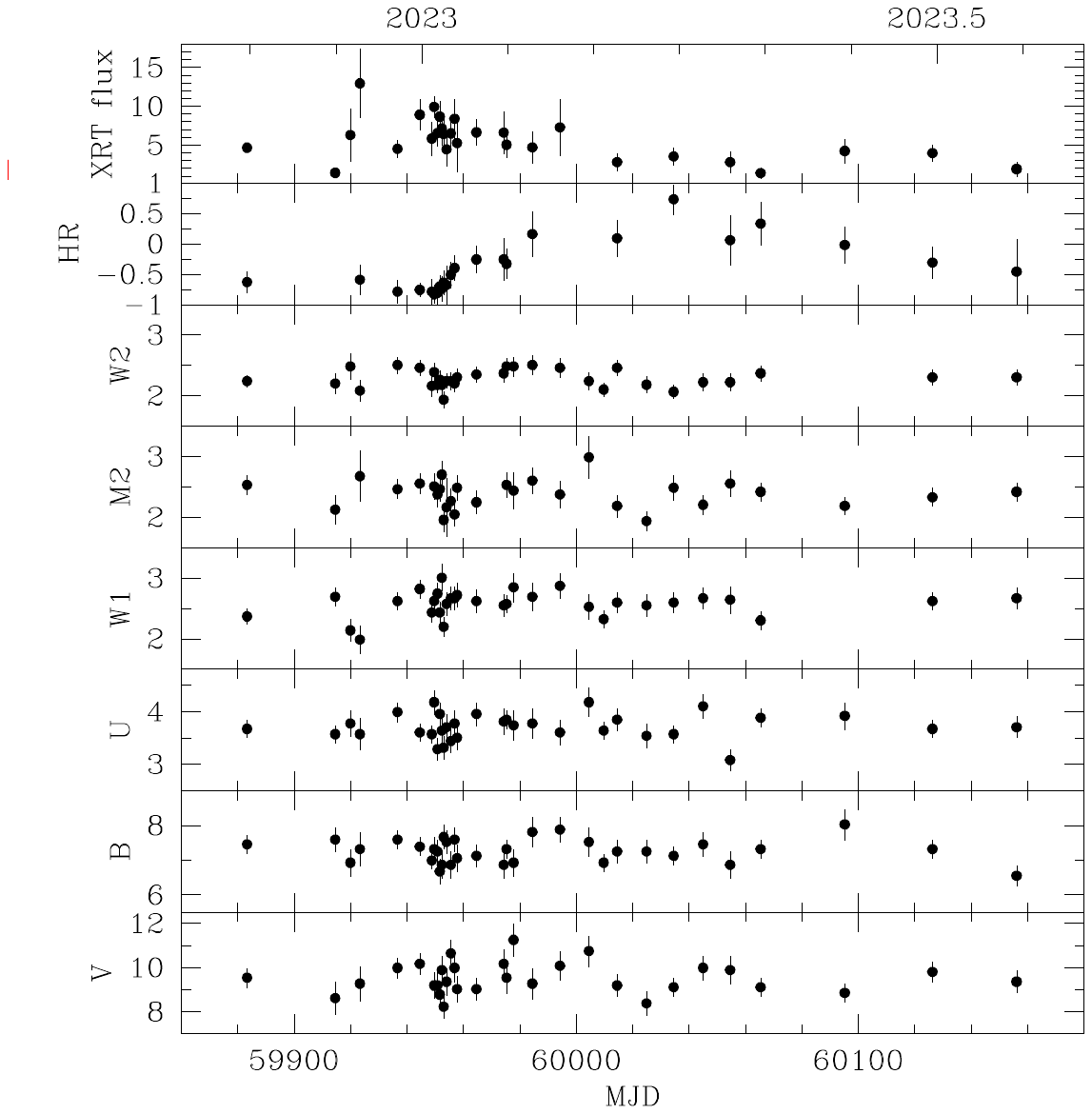}
    \caption{Observed Swift X-ray to optical light curve of IC 3599 from 2022 November to 2023 August. The fluxes in the X-ray and UVOT bands are given in units of $10^{-15}$~ W m$^{-2}$. 
    \label{swift_lc_2023}}
\end{figure*}

Figure\ \ref{swift_lc_2023} shows the \swift\ XRT and UVOT light curves from November 2022 to August 2023. It shows a very dense monitoring phase in January 2023 due to the unusually high count rate of 0.03 counts s$^{-1}$ seen in December 2022. This is a factor of 10 higher than typically observed when it is in its very low state (see also Figure\,\ref{swift_lc}). The light curve also displays a remarkable hardening of the X-ray spectrum after the mini flare
from being very soft at almost HR=-1 to becoming harder with HR$>$0. However, from about March 2023 to the end of the observing campaign in August 2023, IC 3599 has become softer again.

\subsubsection{Variability Analysis}

We have monitored IC 3599 regularly since 2013 which has led to a total number of more than 100 observations. 
This rich data set allows for a detailed temporal analysis. 
As the first step we can apply the Fractional Excess Variance $F_{\rm var} = \frac{\sqrt{\sigma^2 -\delta^2}}{<f>}$ as defined by \cite{rodriguez1997} with an uncertainty of $\sigma_{F_{var}} = \frac{1}{F_{\rm var}} \sqrt{\frac{1}{2N}} \frac{\sigma^2}{<f>^2}$ following the definition by \cite{edelson2002}. Here, $\sigma$ is the variance of the light curve, $\delta$ is the mean value of the uncertainties of the fluxes/count rates, and $<f>$ is the mean value of the fluxes/count rates.  
We applied this to the \swift\ XRT and UVOT light curves including and excluding the 2010 outburst data. The results of the fractional excess variance are listed in Table\,\ref{f_var}.

\begin{table}
 \caption{ Fractional excess variance and the mean fluxes $<f>$ of the \swift\ XRT and UVOT light curves including all data and only the low state 2013-2023 data.  
 \label{f_var}}
\centering
  \begin{tabular}{lcccc}
  \hline
  \hline
& \multicolumn{2}{c}{All Data} & \multicolumn{2}{c}{2013-2022 low state} \\
\rb{Band} & $F_{var}$ & $<f>$ & $F_{var}$ & $<f>$\\ 
\hline\hline
$F_X$ & 2.616\plm0.159 & 3.911 & 0.759\plm0.055 & 2.735 \\
UVW2 & 0.140\plm0.010 & 2.451 & 0.048\plm0.006 & 2.415 \\
UVM2 & 0.156\plm0.011 & 2.531 & 0.037\plm0.010 & 2.482 \\
UVW1 & 0.131\plm0.010 & 2.691 & 0.018\plm0.013 & 2.656 \\
U & 0.112\plm0.008 & 3.786 & 0.041\plm0.006 & 3.741 \\
B & 0.042\plm0.005 & 7.302 & 0.007\plm0.015 & 7.276 \\
V & 0.119\plm0.009 & 9.419 & 0.114\plm0.008 & 9.373 \\
\hline
\hline
\end{tabular}
\end{table}

While the results of the fractional excess variance including the 2010 outburst data seem to follow the decrease of $F_{\rm var}$ with increasing wavelength as expected from accretion disc reprocessing models \citep[e.g.][]{cackett2007}, excluding these data shows a  different picture: The same overall trend is still visible, however 2 bands (U and V) deviate from the systematic trend.

\begin{figure*}
\includegraphics[width=7.8cm]{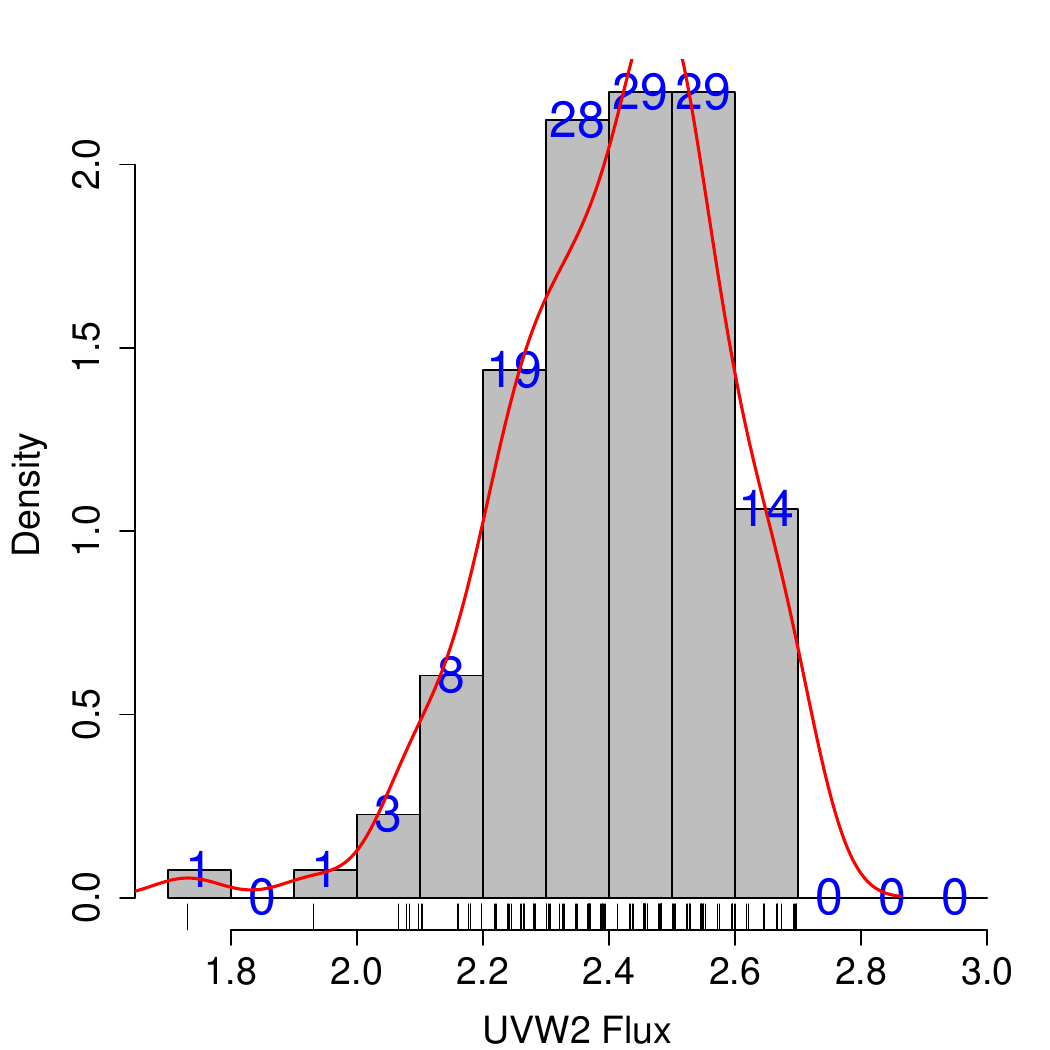}
\includegraphics[width=7.8cm]{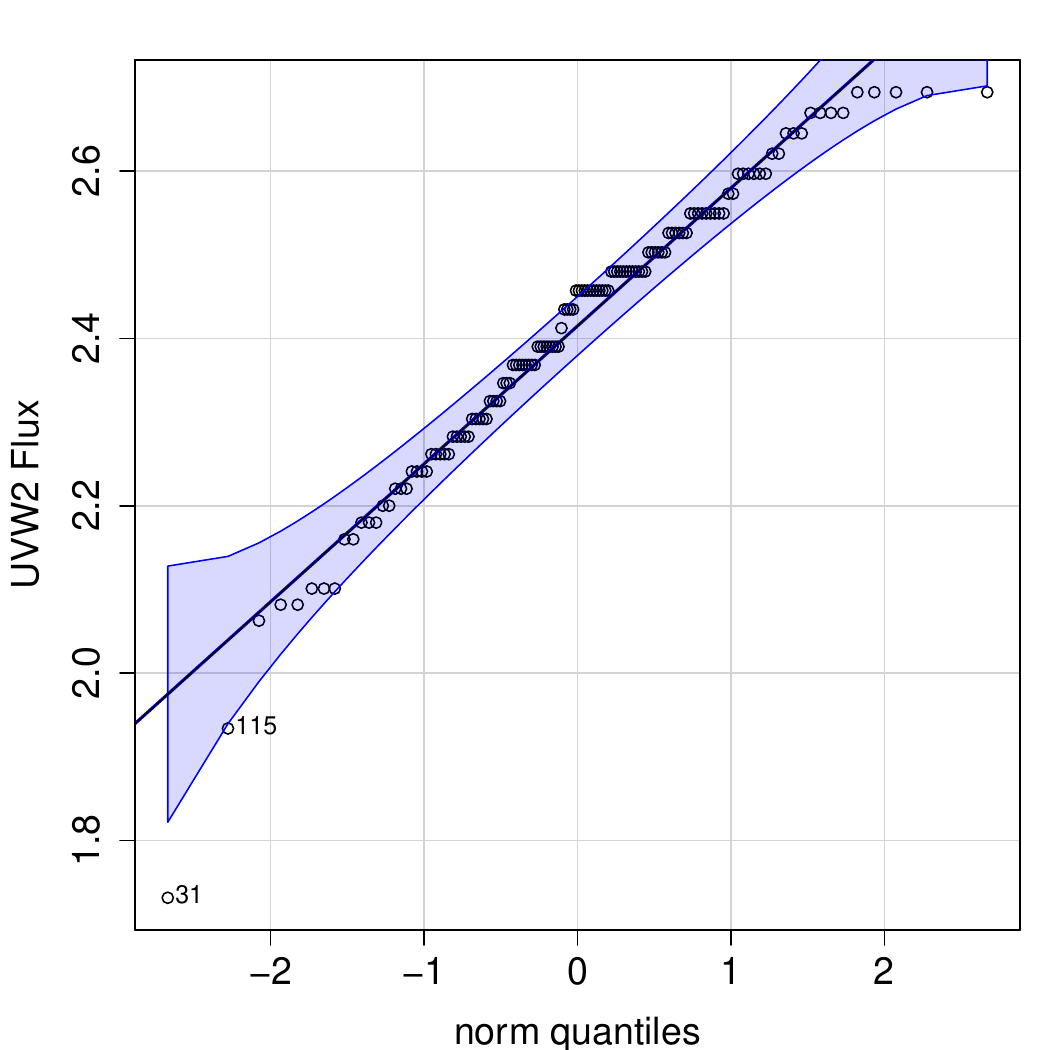}
    \caption{Histogram and QQ-Plot of the W2 fluxes of IC 3599  (in unites of $10^{-15} W~ m^{-2}$) \label{distr_w2}}
\end{figure*}

When working with a variance and mean, the assumption is that the underlying distribution is Gaussian \citep{gauss1821}. We checked the flux distributions in X-rays and in each  UVOT filter using R \citep[e.g., ][]{crawley2009} by looking at the histogram of the flux distribution as well as at the quantile-quantile plot  (QQ plot).
We found that this assumption is well-justified for the UVOT observations. Except for the high state data from 2010, the data of all UVOT observations are well-described by a Gaussian distribution 
in all 6 filters.  Figure\,\ref{distr_w2} displays the distribution and the QQ plot of the W2 magnitudes of IC 3599 which clearly shows that this is a Gaussian distribution. The other UVOT filters show similar distributions which are all consistent with a Gaussian distribution. 

\begin{figure*}
\includegraphics[width=7.8cm]{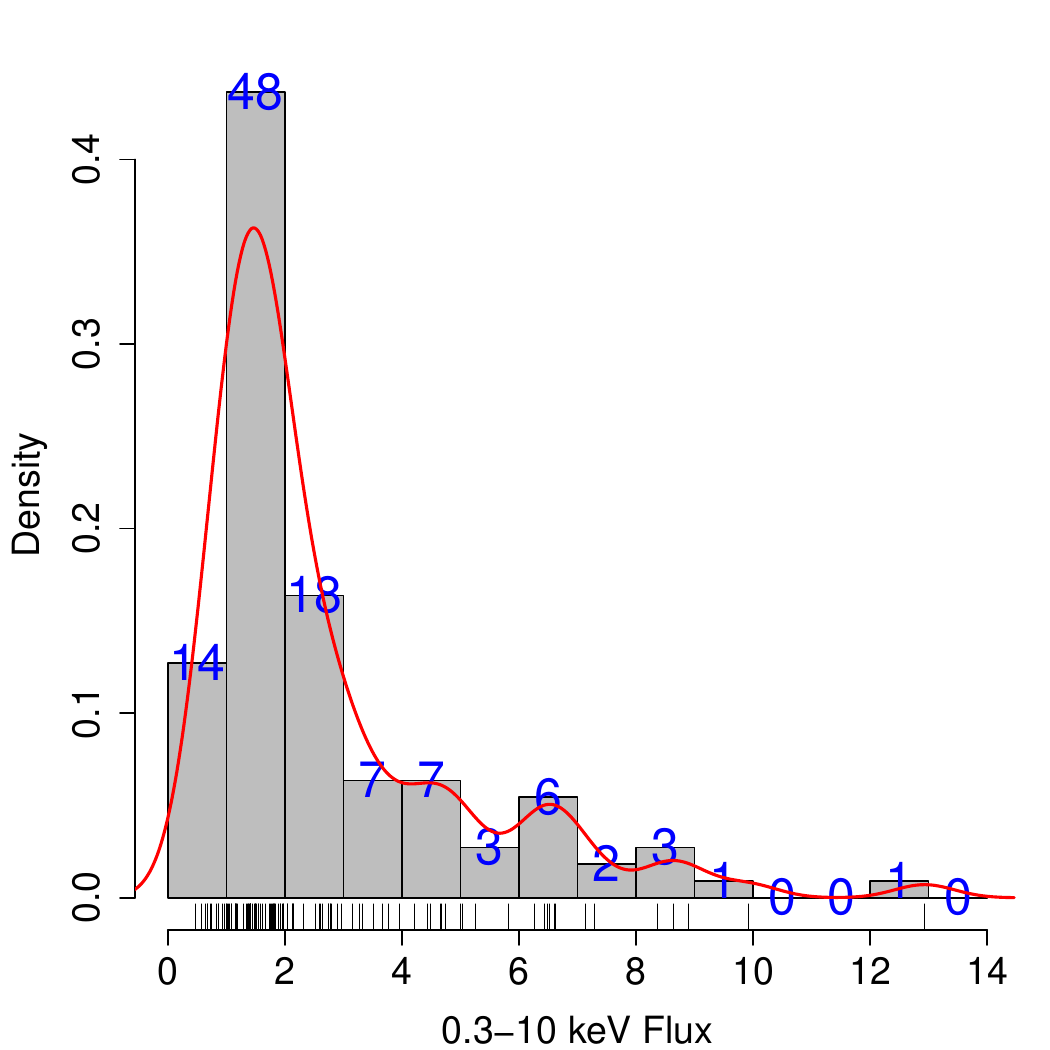}
\includegraphics[width=7.8cm]{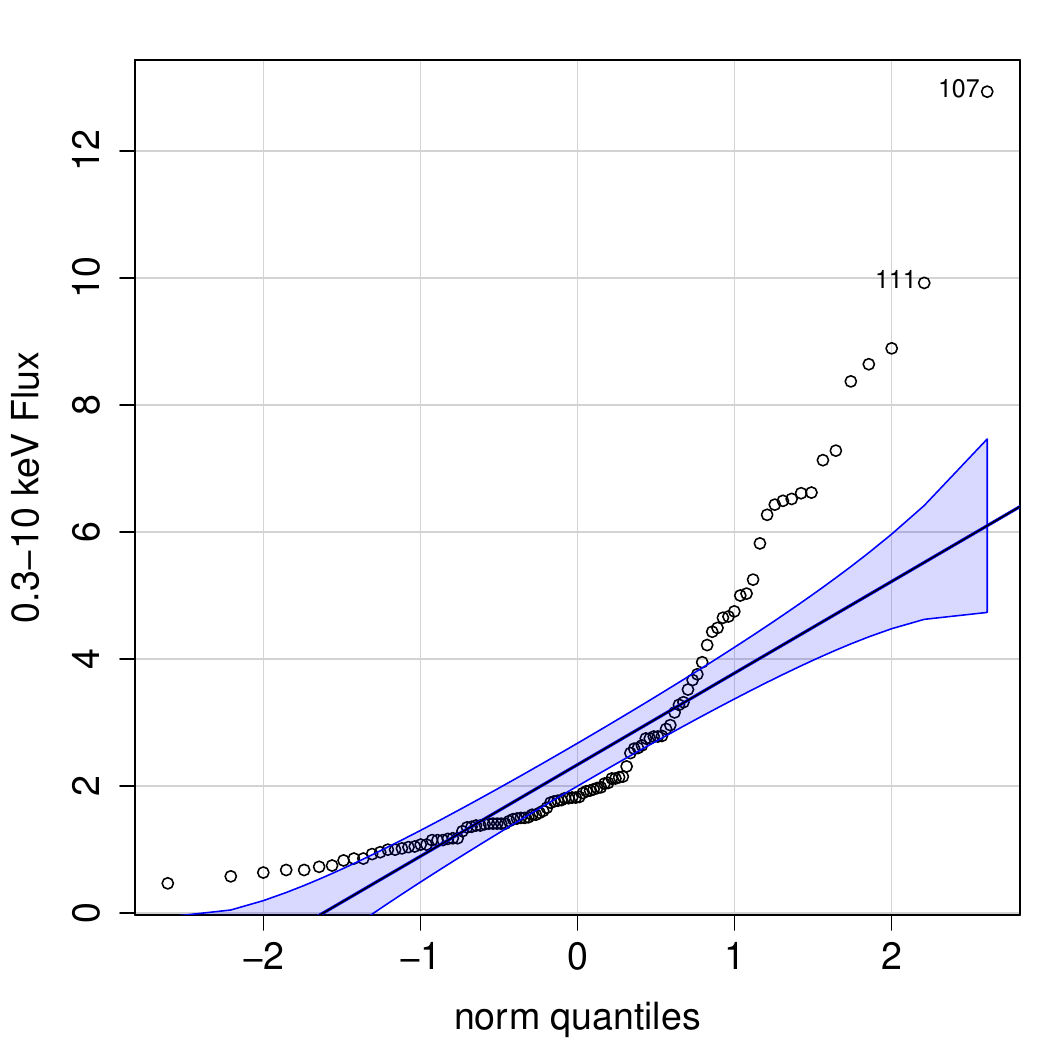}
    \caption{Histogram and QQ-Plot of the 0.3-10 keV X-ray fluxes of IC 3599 (in units of $10^{-16} W~ m^{-2}$) \label{distr_fx}}
\end{figure*}

However, the X-ray flux variations are not represented by a Gaussian distribution. This is clearly shown in Figure\,\ref{distr_fx} which displays the histogram and the QQ-plot. 

For non-Gaussian distributions, the variance and the mean are not good measures of a sample and in these cases, the median is a much better measure than the mean. In order to study the overall variability in all Swift XRT and UVOT light curves we designed a new measure:  a variability parameter 

\begin{equation}
P_{\rm var} = \frac{1}{\sqrt{N}} \frac{1}{x_{\rm med}} \sqrt{\sum_i{(x_i - x_{\rm med})^2}\times \frac{\delta_{med}^2}{\delta_i^2}}
\end{equation}

which is like a reduced $\chi^2$ value. This parameter, $P_{\rm var}$, is a measure of how strongly a source deviates from  the median value. Here $x_{\rm med}$ and $\delta_{\rm med}$ are the sample median values of the fluxes and the median of the uncertainties. The values $x_i$ and $\delta_i$ are the individual values with their uncertainties. The term $\frac{\delta_{med}^2}{\delta_i^2}$ is a weighing factor that takes the individual uncertainties into account so that values with large uncertainties carry less weight in the calculations than those values with small uncertainties. 
We estimate the uncertainties on the values of $P_{\rm var}$ by $\Delta P_{\rm var} = \frac{\delta_{med}}{x_{med}} P_{\rm var}$. The values for $P_{\rm var}$ are listed in Table\,\ref{p_var}. These values also confirm the stronger variability in X-rays than in the UV and optical. 
The variability parameter $P_{\rm var}$ also shows that these values are not dominated by the 2010 outburst measurements. It allows a direct comparison with the variability strength between Gaussian distributed data and non-Gaussian data, like the X-ray flux distribution of IC 3599, simply because it does not assume a Gaussian distribution and it is therefore a more general measure of the variability. 

\begin{table}
 \caption{Variability parameter $P_{\rm var}$ and the median flux $f_{med}$
 of the \swift\ XRT and UVOT light curves of the low state 2013-2023 data
 \label{p_var}}
\centering
  \begin{tabular}{lcccc}
  \hline
  \hline
& \multicolumn{2}{c}{all data} & \multicolumn{2}{c}{2013-2022 low state} \\
\rb{band} & $P_{var}$ & $f_{med}$ & $P_{var}$ & $f_{med}$\\ 
\hline\hline
$F_X$ & 1.072\plm0.410 & 1.830 & 0.808\plm0.306 & 1.820 \\
UVW2 & 0.107\plm0.006 & 2.457 & 0.073\plm0.004 & 2.457  \\
UVM2 & 0.112\plm0.007 & 2.508 & 0.073\plm0.005 & 2.508  \\
UVW1 & 0.107\plm0.006 & 2.667 & 0.059\plm0.003 & 2.667 \\
U & 0.093\plm0.004 & 3.772 & 0.060\plm0.003 & 3.772  \\
B & 0.049\plm0.002 & 7.318 & 0.039\plm0.002 & 7.319  \\
V & 0.094\plm0.005 & 9.529 & 0.088\plm0.004 & 9.529 \\
\hline
\hline
\end{tabular}
\end{table}

Just a Gaussian flux distribution by itself does not prove that these are just random fluctuations. In order to check on any variability we performed a periodogram within R using {\it{spec.pgram}}. Besides some red noise at lower frequencies, we only found white noise in the data. The V filter data are dominated by white noise.
This variability pattern is typical for type 2 AGN \citep{wang2024} and may suggest that IC 3599 remains in its Seyfert 1.9 state at this time. 

\newpage

\subsubsection{Merged XRT and UVOT Light Curves \label{merged}}

\begin{figure*}
\includegraphics[trim=10 150 15 280,clip,width=\textwidth]{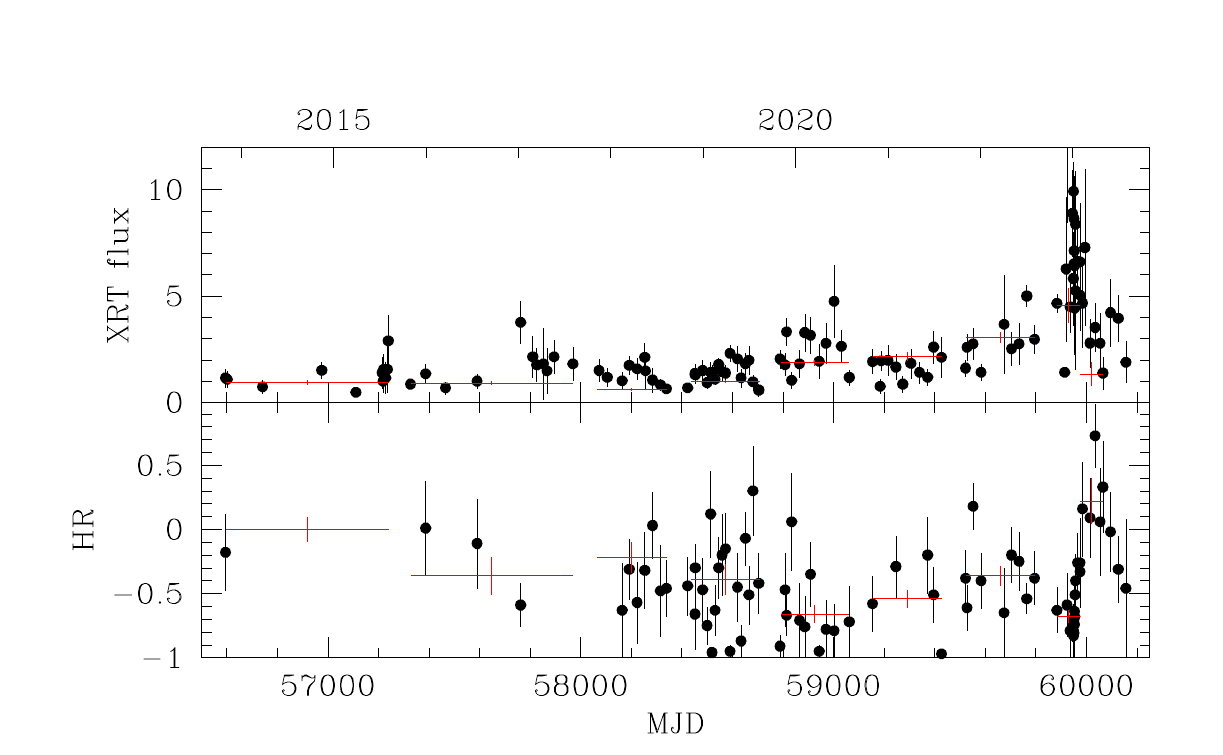}
    \caption{Swift XRT 0.3-10 keV flux (in units of $10^{-16}$ W m$^{-2}$) and hardness ratio long-term light curves. The black data points displayed in Figure \ref{swift_lc} show detections. The red data points display merged data.  \label{swift_xrt_lc}}
\end{figure*}

Figure\,\ref{swift_xrt_lc} displays the long-term Swift XRT light curve with the data averaged over each year, except for the 2013-2016 period because the number of observations is relatively small and the count rate was low, of the order of 0.003 counts s$^{-1}$. We binned the 2013-2015 and 2015-2017 observations into one bin each and after 2017, the observations of each year were combined. The binned data are displayed as red data points in Figure\,\ref{swift_xrt_lc}. The 0.3-10 keV fluxes and the hardness ratios for these periods are listed in Table\,\ref{xrt_merge}.  There is clearly an overall slow brightening in X-rays over the last decade. The hardness ratio light curve also suggests spectral variability with IC 3599 becoming softer from 2013 to 2020, but  becoming harder again since. However, there is no correlation between the hardness ratio and the X-ray brightness of IC 3599.

\begin{figure*}
\includegraphics[trim=10 150 15 15,clip,width=\textwidth]{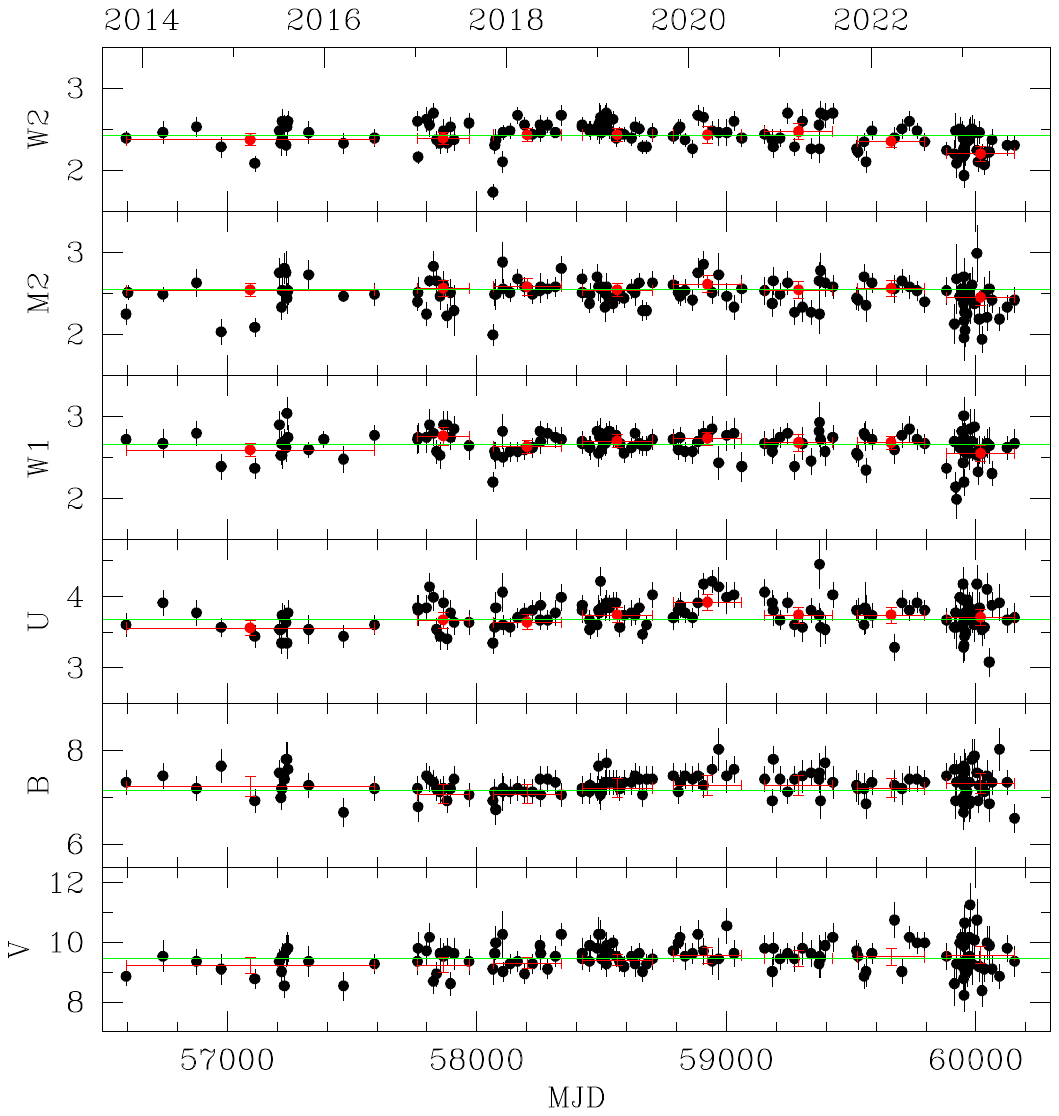}
    \caption{Swift UVOT long-term light curves.  The fluxes are given in units of $10^{-15}$ W m$^{-2} $. The black data points display as in Figure \ref{swift_lc} each observation with a detection. The red data points display merged data as listed in Table\,\ref{uvot_merge}.  The vertical green lines mark the median values in each of the filters. \label{swift_uvot_lc}}
\end{figure*}

Figure\,\ref{swift_uvot_lc} displays the UVOT light curves between 2013 and 2023 with the merged data shown in red. Overall, there is no significant variability from year to year. The results for the merged data are listed in Table\,\ref{uvot_merge}. Figure\,\ref{swift_uvot_lc} and Table\,\ref{uvot_merge} show  that there is no significant variability on long or short time scales, not even a trend like seen in X-rays. This is somewhat expected given the Gaussian distribution of the UVOT data.

\subsection{X-ray Spectral Analysis}
For the comparison of the high and low states we merged the 2010 February and May data into a high state spectrum and did the same for the low state by merging all data from 2013 to 2023. The high and low state data were then fitted with various spectral models. The results of these fits are listed in Table\,\ref{swift_xspec_fit}. All models were fitted with the z=0 absorption fixed to the Galactic value of $1.17\times 10^{20} cm^{-2}$. 

\begin{figure}
\includegraphics[width=9.5cm]{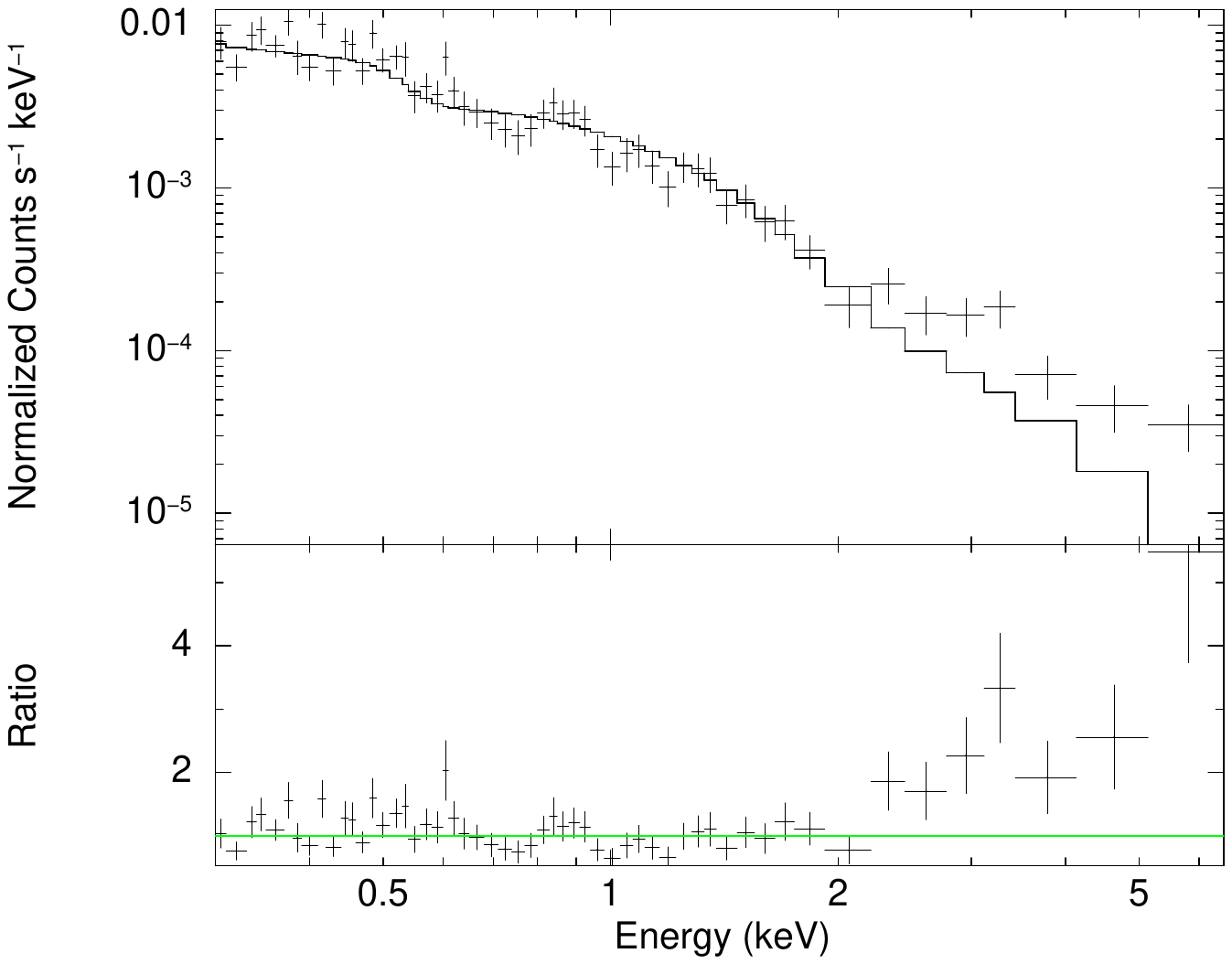}
    \caption{Upper panel: Swift XRT spectrum of all low state 2013-2023 data merged fitted by an absorbed single power law model. Lower panel: The ratio between the data and the model. 
    \label{swift_xspec}}
\end{figure}

The total exposure time of the high state spectrum was 3354s and the count rate 0.115 counts s$^{-1}$ which means a total number of source counts of 260 counts which account for 99.3\% of the counts in the spectrum. 
The relatively low number of counts in the high-state spectrum only allowed for an analysis using W statistics \citep{cash1979}.

 The overall count rate of the low-state spectrum is 0.0046 counts s$^{-1}$ and has a total exposure time of 302 ks. This results in a total number of about 1400 source counts, which accounts for 89\% of the counts in the spectrum. The relatively large number of counts allows us to fit the low-state spectrum with  $\chi^2$ statistics. The binning of the low-state spectrum was 20 counts per bin.  This spectrum is displayed in Figure\,\ref{swift_xspec}.

The high and low flux state spectra were first fitted by a single power law model. 
While in the high state case the data can be  fitted by this model acceptably,  the low state spectrum clearly deviates from a simple power law model as shown from the fit parameters in Table\,\ref{swift_xspec_fit} and in Figure\,\ref{swift_xspec}. One thing to keep in mind is that the number of counts in the high state data above 1.5 keV is so low that we do not have any information on the spectrum above this energy. Next, we applied a broken power law model and a blackbody plus power law model. These are phenomenological models that can describe the spectra quite well in both cases. Finally, we used a neutral partial covering absorber model. For the high state data, the parameters of the partial covering absorber are unconstrained. There are no data at higher energies that are needed to fit this model. For the low state data however, the spectrum is well described by a partial covering absorber model with a column density of $6\times 10^{22} cm^{-2}$ and a covering fraction of 80\%.

\begin{table}
 \caption{ Spectral analysis of the high and low state XRT spectra. The high state data were fitted with W statistics, while for the low state data, $\chi^2$ statistics was applied. \label{swift_xspec_fit}}
\centering
  \begin{tabular}{lcc}
  \hline
  \hline
Parameter & 2010 High  & 2013-2023 Low  \\
\hline
\multicolumn{3}{c}{powerlaw} \\
\hline
$\Gamma$ & 4.32\plm0.20 & 3.12\plm0.10 \\
$\chi^2/\nu$  & 111/82 & 87.9/53 \\
\hline\hline 
\multicolumn{3}{c}{broken power law} \\
\hline
$\Gamma_1$ & 3.51\plm0.40 & 3.29\plm0.15 \\
$E_{\rm break}$ [keV] & 0.85\plm0.13 & 1.53$^{+0.62}_{-0.38}$ \\
$\Gamma_2$ & 7.30\plm1.50 & 1.96$^{+0.43}_{-0.67}$ \\
$\chi^2/\nu$ & 76/82 & 61.3/51 \\
\hline\hline
\multicolumn{3}{c}{blackbody + power law} \\
\hline
kT [eV] & 112\plm12 & 95\plm9 \\
$\Gamma$ & 4.56\plm1.65 & 2.18\plm0.26 \\
$F_{\rm bb}$ [W m$^{-2}$] & $3.8\times10^{-15}$ & 7.5$\times 10^{-17}$ \\
$F_{\rm po}$ [W m$^{-2}$] & $2.0\times10^{-15}$ & 9.6$\times10^{-17}$ \\
$\chi^2/\nu$ & 76/82 & 57.9/51 \\
\hline\hline
\multicolumn{3}{c}{xpcfabs $\times$ power law} \\
\hline
$N_H^1$ & 6.30$^2$  & 6.30$^{+4.62}_{-2.51}$ \\
$f_{pc}$ & 0.05$^3$   & 0.80$^{+0.07}_{-0.11}$ \\
$\Gamma$ & 4.32\plm0.20  & 3.27\plm0.12 \\
$\chi^2/\nu$ & 111/82  & 60.1/51 \\
\hline
\hline
\end{tabular}

$^1$ The redshifted (z=0.021) partial covering absorber column density is given in units of $10^{22} cm^{-2}$.
$^2$ The partial covering absorber column density is fixed to $6.3\times 10^{22} cm^{-2}$. 
$^3$ not constraint

\end{table}

\begin{figure*}
\includegraphics[width=8.5cm]{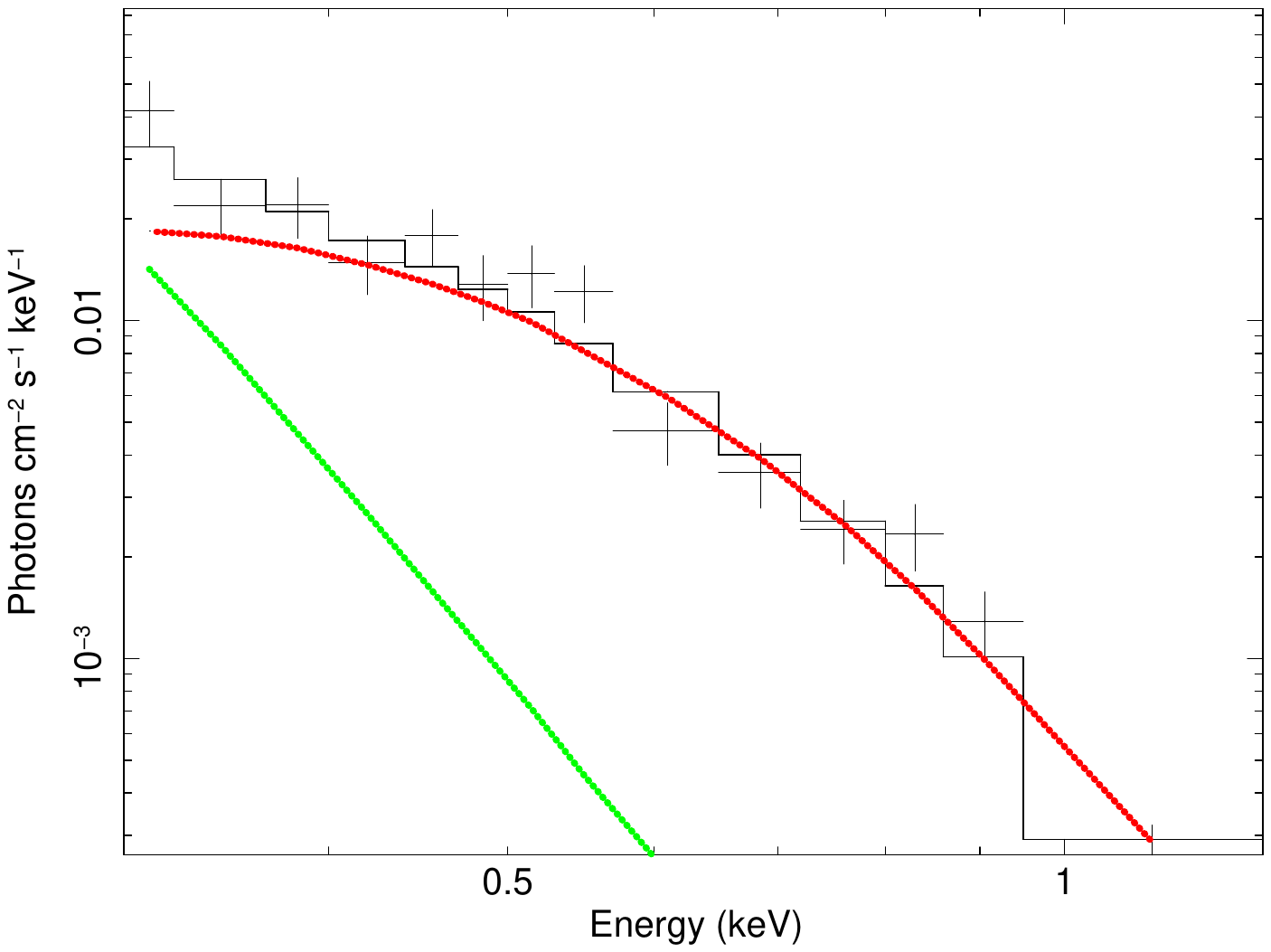}
\includegraphics[width=8.5cm]{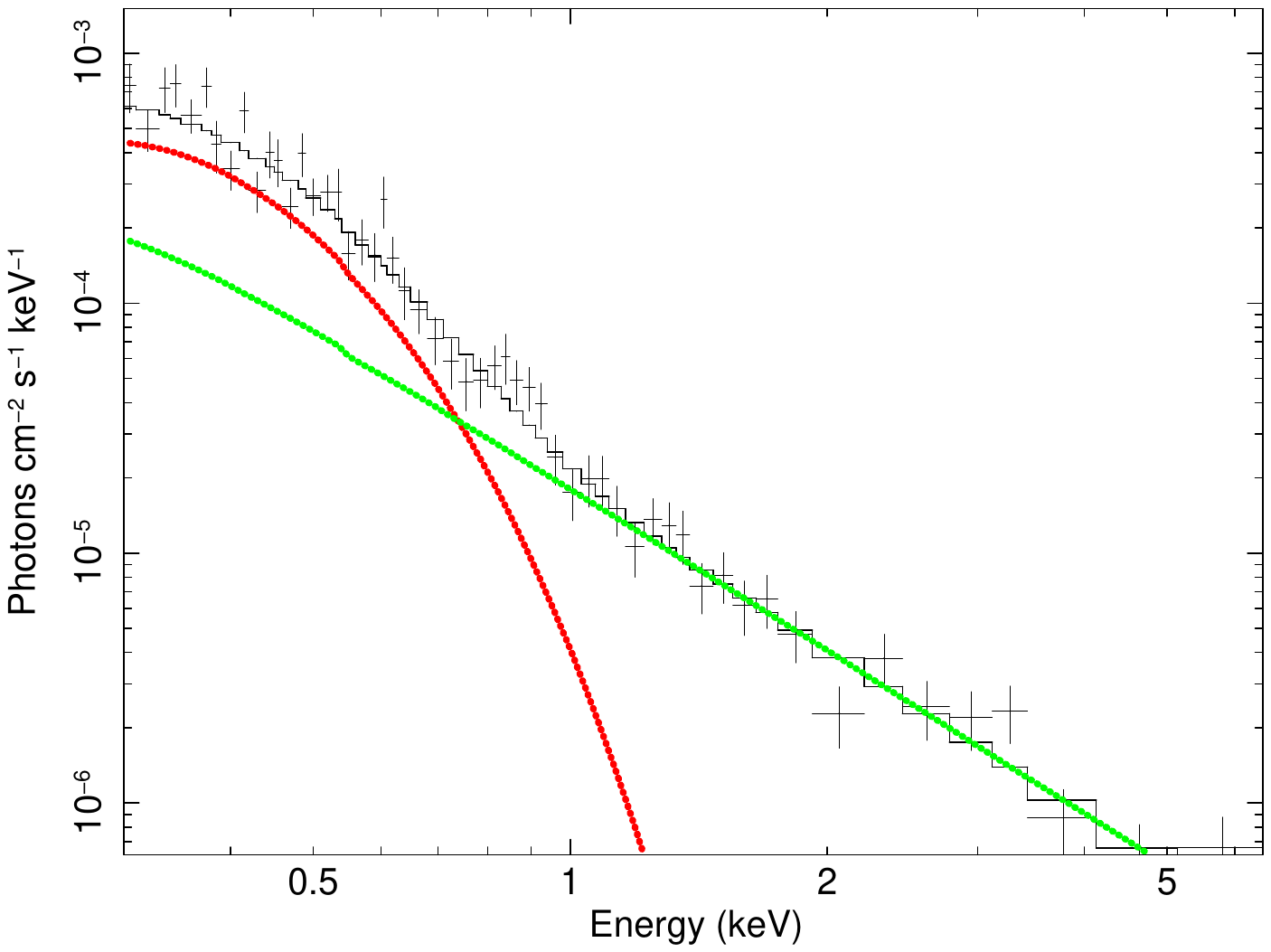}
    \caption{Swift XRT merged high and low state spectra (left and right, respectively) The red line displays the black body spectrum and the green line, the power law component. \label{swift_xspec_high_low}}
\end{figure*}

Figure\,\ref{swift_xspec_high_low} displays an absorbed blackbody plus power law model fit to the merged  high state data from 2010 and the 2013-2022 low state data. Note, that for display purposes the high state data have been binned, however the spectral fits were performed with unbinned data using W-statistics \citep{cash1979}. 
The blackbody component is displayed in red while the power law component is shown in green.  
As for the change between the high and low state black body and power law components, the black body component changes by a factor of 48$\pm$13 and the power law component by a factor of 23$\pm$6. This suggests that the blackbody component decreased stronger than the power law component.

\subsection{Strong X-ray Spectral Change in January 2023}
As shown in the hardness ratio light curve in Figure\,\ref{swift_lc_2023} there is a strong spectral change from being a very soft source to a significantly harder AGN around MJD 59954-59977 (2023-January-10 to 31).  In order to study the change in the X-ray spectrum, we merged the data after IC 3599 emerged from its sun constraint on 2022-October-31  to 2023-January-10 when it was in the soft state and between 2023-February-10 and May-01 when it was in the hard state. Both spectra can be fitted by simple power law models. For the soft and hard phases, the photon indices were $\Gamma = 4.03^{+0.37}_{-0.33}$ and $\Gamma = 1.70^{+0.87}_{-0.54}$, respectively. During the soft state the AGN is also brighter with a flux of $F_{0.3-10 keV} = (4.33\pm0.40)\times 10^{-16} W m^{-2}$, while during the hard state the flux was at $(1.34\pm0.58)\times 10^{-16} W m^{-2}$. Since about March 2023 IC 3599 has transitioned again into a softer X-ray state.

\subsection{UV Morphology}

\begin{figure}
\includegraphics[trim=100 280 120 240, clip, width=8.5cm]{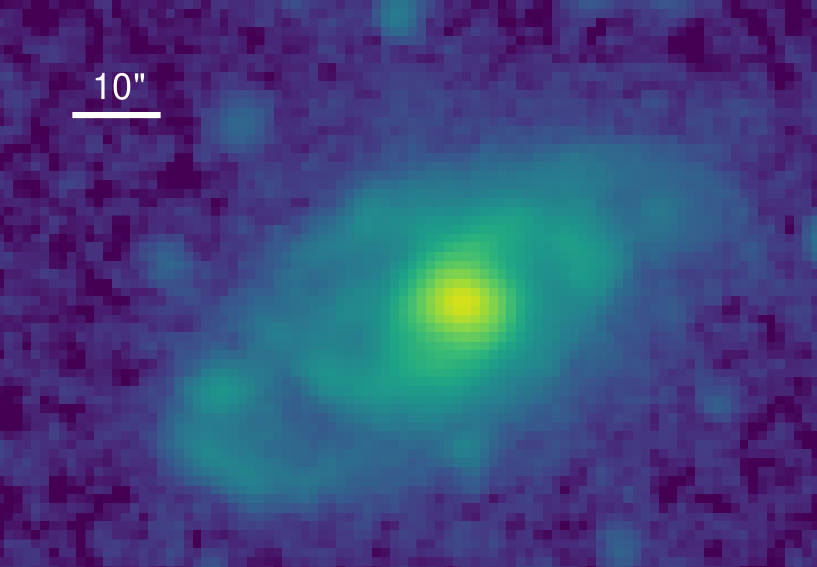}
    \caption{Swift UVOT  W2 image of IC 3599. At the distance of IC 3599, 10" corresponds to 4.4 kpc. \label{swift_image}}
\end{figure}

Figure\,\ref{swift_image} displays the W2 image which was merged from all low-state W2 data from 2013-2023. This image has a total exposure time of 99 ks.  The image clearly shows the spiral structure of the host galaxy,  tracing the young, blue, stars in the spiral arms. The projected angular diameter of the galaxy is about 75'' which corresponds to a physical size of 33 kpc; larger than a typical dwarf galaxy, suggesting that the SMBH mass of IC 3599 is not particularly low.

\subsection{UV Colors}

Although the Gaussian distributions of the magnitudes in all 6 UVOT filters during the low state suggest random fluctuations, we still checked if there may be any dependence on the colors with fluxes in X-rays and the UV. For this purpose, we calculated U-B and W2-W1 colors and correlated them with various properties, like W2 magnitude, count rate and hardness ratio. As expected, we did not find any correlations among these parameters.

\subsection{Spectral Energy Distribution}

\begin{figure}
\includegraphics[width=6.5cm, angle=270]{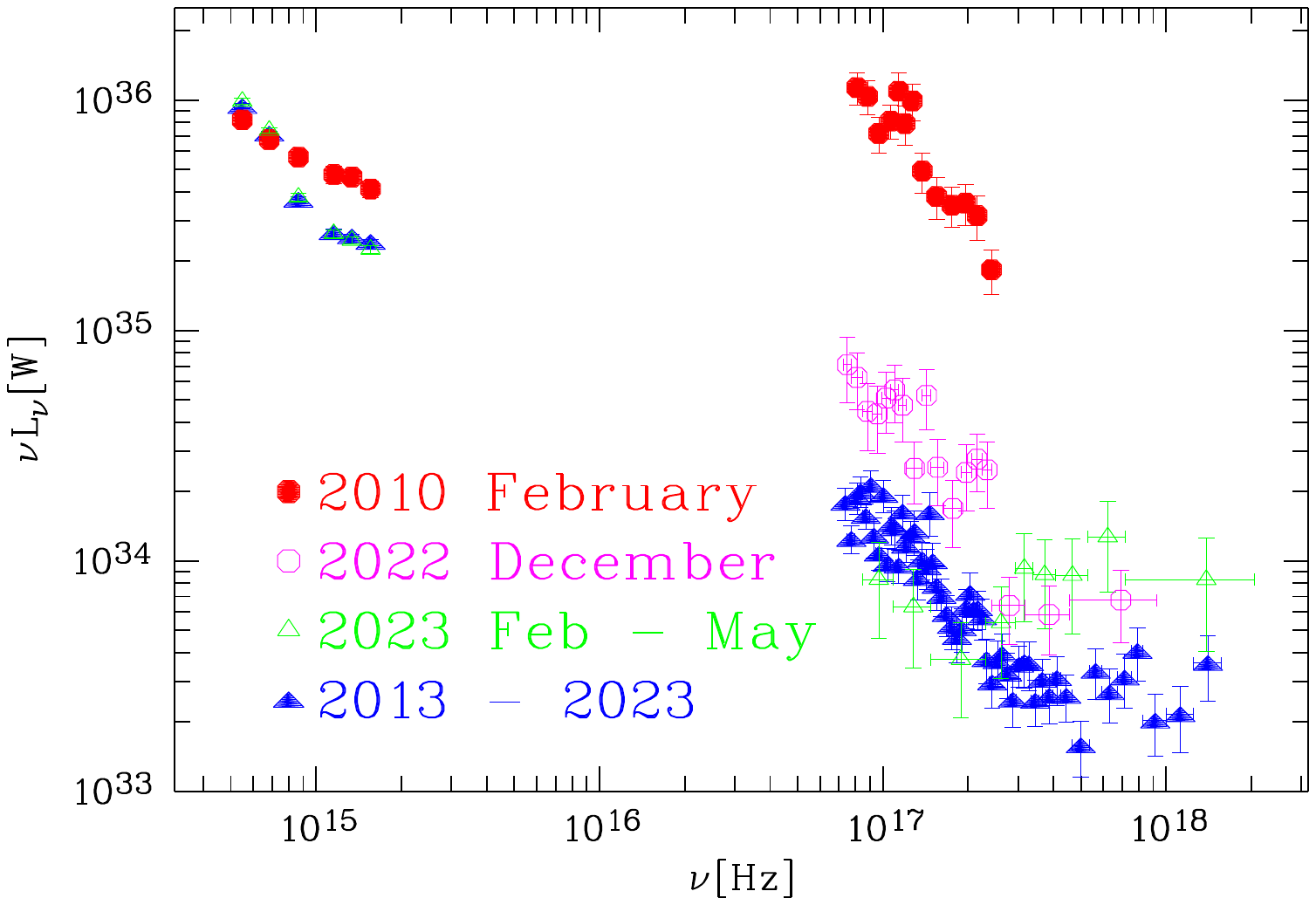}
    \caption{The SED of IC 3599 between the outburst in February 2010 (red), the  merged 2013-2023 low state data (blue), the mini flare data in December 2022 (magenta), and the merged data after the mini flare (green). Note that the UVOT data during this flare are not shown due to their similarity to the overall low-state data. \label{ic3599_sed}}
\end{figure}

The Spectral Energy Distribution (SED) of IC 3599 during the February 2010 outburst and the 2013-2023 low state is shown in Figure\,\ref{ic3599_sed}. It displays how dramatically different the SED of IC 3599 was during the 2010 outburst. The optical/UV to X-ray spectral slope \aox\footnote{The optical/UV to X-ray spectral slope \aox\ is defined by \cite{tananbaum1979} as \aox = $-0.384\times (log(l_{\rm 2keV}) - log(l_{2500\AA}))$, where $l_{\rm 2500\AA}$ and $l_{\rm 2keV}$ are the luminosity densities at 2500\AA\, and 2 keV.} during the outburst was \aox=1.57\plm0.07 and during the low state \aox=1.79\plm0.04. Note that the uncertainties on the \aox\ value during the outburst maybe larger, because to determine the 2 keV data point we assumed a simple power law model, which may not be the correct model at 2 keV, but with the lack of available data it is the value derived from the X-ray spectral fit. 

For the 2010 outburst data and applying the simple power law model, we derived a 0.2-2.0 keV luminosity of $L_X = (9.9\pm0.8)\times 10^{35}W$. Applying the relation $log L_{\rm bol} = 1.23\times log(L_X) -7.36$ as given in \cite{grupe2010} we estimated the bolometric luminosity  as $8\times 10^{36}$ W. Assuming a black hole mass of IC 3599 of 2.5$\times 10^6 M_\odot$ the Eddington luminosity is $3.15\times 10^{37}W$ which then results in an Eddington ratio of $L/L_{\rm Edd}$ = 0.25.

As for the low state 2013-2023 data, assuming the broken power law model as shown in Table\,\ref{swift_xspec_fit}, the 0.2 - 2.0 keV luminosity is $1.8\times 10^{34} W$ which results in a bolometric luminosity of $6\times 10^{34} W$ and an Eddington ratio of $L/L_{\rm Edd} = 2\times 10^{-3}$. 

In addition, the SED (Figure\,\ref{ic3599_sed}) also displays the SED of the data during the December 2022 flare period (open magenta circles) and the data after the mini flare from February to May 2023 (open green triangles). 
Note that for visualization purposes we have binned the data with a binning of 10. For the spectral fits, no binned data were used applying W statistics \citep{cash1979}.
The X-ray data after the mini flare
indicate a significant flux increase in the hard X-ray band. The spectral analysis of these data results in a very flat X-ray spectrum with $\Gamma = 1.70^{+0.81}_{-0.51}$, significantly flatter than what has been observed during the entire 2013-2022 low state period. Note that the low number of counts did not allow for a more sophisticated spectral analysis.

\section{Discussion}

\subsection{High-Amplitude Variability of AGN}

In principle, any single one of a huge number of small or large flares in any blazar or radio-quiet AGN could be a TDE. However, TDEs are very rare events, and almost every single AGN or blazar is known to vary \citep[see][for reviews]{ulrich1997, gaskell2008, gallo2018}. Therefore, exceptional positive evidence is required for any claim of a TDE in an AGN, as emphasized by, e.g., \citet[][]{rees1988} and \citet{komossa1999a}. The situation is completely different, when a {\em{quiescent}} galaxy suddenly shows a huge X-ray outburst, because the only known model to explain a quasar-luminosity outburst in a galaxy without long-lived accretion disk is a TDE. 
IC 3599 is a bona fide AGN with a long-lived NLR \citep{komossa1999a, grupe2015}, and closely resembles the increasing number of changing-look AGN identified in recent years \citep[e.g.][]{Alloin1986, runco2016, macLeod2019, frederick2019, ochmann2024}. 

Therefore, so far, no definite positive evidence for a rare TDE in this highly variable AGN has been presented. However, we note that \citet{Mandel2015} mentioned in passing the possibility of tidal disruption of one star and later tidal stripping of the second star of a binary star system in context of IC 3599. In that case, the near-identical peak luminosities of the two outbursts of IC 3599 would be unexpected. 
Further, w.r.t. the two outbursts of IC 3599, \citet{komossa2014} pointed out that during certain phases of binary SMBH evolution, or in the presence of a recoiling SMBH, TDE rates can be temporarily strongly enhanced \citep{chen2009, KomossaMerritt2008, stone2011}, but two events within decades would still be rare, and IC 3599 does not show evidence for a recent major merger (e.g., Figure \ref{swift_image}), and the near-identical peak luminosities of the two outbursts would have to be coincidence. Therefore, \citet{komossa2014} re-emphasized that TDEs are most reliably identified in quiescent galaxies. 
Further, \citet{grupe2015}
commented on the TDE model of \citet{liu2009} \citep[see also][]{liu2014} in application to IC 3599. \citet{liu2009} predicted TDE light curves for stellar disruption in a binary SMBH system, showing characteristic recurrent flare and dip events. However, the timing and amplitudes of IC 3599 did not fit that model. {\footnote{Even though not discussed further here, we note that multiple outburst events in {\em{quiescent}} galaxies have been interpreted as possible partial stripping events; for example the case of RXJ133157.6324319.7 \citep{hampel2022} which showed a second outburst \citep{malyali2023} $\sim$30 yrs after the first. }}

The one TDE model with explicit predictions which  we can now  test and rule out with the new monitoring data presented here, is the published repeat-tidal-stripping scenario for IC 3599, because that model made testable predictions for the time frame covered in our new observations. We come back to this model in the following paragraphs. 
Beyond that, we continue to consider AGN scenarios, rather than TDE scenarios, for this highly variable AGN. 
Below, we also discuss the other aspects of the long-term variability of IC 3599, unrelated to the two big outbursts.

\subsection{No Outburst of IC 3599 in 2019/2020}

\citet{campana2015} predicted that there would be another outburst due to the partial tidal stripping of an orbiting star in 2019 and 2020. They predicted a 9.5 year period. The model also assumed a black hole mass of the order of a few $10^5 M_{\odot}$. While in general a TDE scenario cannot be excluded in principle, as pointed out by \citet{grupe2015} and \citet{komossa1999a}, IC 3599 is an AGN and the dramatic flaring in IC 3599 does not have to be associated with a tidal disruption event and that given the 
black hole mass estimates of IC 3599 of the order of several millions of solar masses, an accretion disk instability scenario would be more likely. \citet{campana2015}'s prediction of repeat flaring made it possible to test the tidal stripping scenario. The {\it Swift} monitoring programs from 2019 and 2020 initiated by us and Campana clearly show that there is no outburst seen during this period (see Figure\,\ref{swift_lc}). This result clearly favors the accretion disk scenario as suggested by \citet{grupe2015}.

\subsection{Slow Increase in X-ray Flux}

Figure\,\ref{swift_xrt_lc} suggests that there has been a slow overall increase in X-ray flux since about 2018. In the accretion disk scenario that could be explained by a slow fill-up of the inner part of the accretion disk which emptied out during the last outburst in 2010.   In \citet{grupe2015} we estimated the fill-up time to be several decades, following the relation $\tau_{\rm fill} = 0.33 \alpha^{-8/10} M_6^{6/5} \dot{M}^{-3/10}_{\rm edd} (R_{\rm trunc}^{5/4} - R_0^{5/4})$ given in units of month
following the relation by \citet{saxton2015}. The other parameters are the viscosity $\alpha$ for which we assumed a standard value of 0.1, the black hole mass $M_6$ in units of $10^6 M_{\odot}$, the infill rate $\dot{M_{\rm edd}}$ in units of the Eddington limit. The truncation radius $R_{\rm trunc}$ and the inner radius $R_0$ are given in units of the gravitational radius $R_g = \frac{GM}{c^2}$. 

While the X-ray flux seems to develop some long-term trend to become slightly brighter, there is no sign of any significant variability in the optical and UV. IC 3599 seems to be fairly constant in all 6 UVOT filters and the fluctuations seem to be random, which is supported by the Gaussian distributions of the magnitudes seen in each filter. 

\subsection{X-ray Mini Flare and Spectral Hardening Event in 2022/2023}

As shown in Figure\,\ref{swift_xrt_lc}, there is an X-ray mini flare at the beginning of December 2022. Right after this short-lasting flare we see a dramatic hardening of the X-ray spectrum, as shown in the hardness ratio light curve in Figure\,\ref{swift_xrt_lc}. This hardening of the spectrum can also be seen in the spectral energy distribution in Figure\,\ref{ic3599_sed}. The data from this hardening event as displayed as green open triangles. 

The mini flare started on 2022 December 01 and peaked on December 10 with a peak flux of 1.29\plm0.45 $\times 10^{-15} W~m^{-2}$, which is a factor of about 10 higher than the flux on December 01 (see also Table\,\ref{swift_res}). 
While the spectrum was very soft during the flare, the onset of the spectral hardening started around 2023 January 03 and reached the hardest state with a hardness ratio of 0.73\plm0.25 on March 31. 
The spectrum became soft again by the end of May 2023 and has remained soft in all observations after that. 

The question is raised, which physical process(es) cause this temporary spectral hardening. The hard X-ray spectral component, typically observed in all AGN, is usually interpreted as the presence of an accretion-disk corona.
It is therefore possible that a corona formed temporarily, following the mini flare, and disappeared again in the low-state. 
Alternatively, we might speculate that a temporary absorption event caused an apparent spectral hardening. For instance, it is possible that due to the flaring activity a temporary outflow was triggered because of the increase in the accretion luminosity \citep{takeuchi2013}. When a clumpy outflow partially covers the intrinsic emission, a spectral hardening can result. However, even at the peak of the mini flare, $L/L_{\rm Edd}$ is only about 0.02, whereas launching of radiation-pressure driven outflows requires values closer to 1. 

The Swift snapshot observations do not contain enough photon statistics to distinguish between these different spectral models (like emission  from a temporary corona, versus partial covering absorption, or more complex spectral models involving the interplay of several components). Deeper, triggered, follow-up observations during the flare decline phase will be very useful, if new mini flares are detected in the future.

\subsection{Black Hole Mass Estimates}

While the repeating TDE model of \citet{campana2015} requires a small black hole mass of a few $10^5 M_{\odot}$, all observational parameters, like the width of the [OIII]$\lambda$5007 line, suggest a black hole mass that is at least 10 times lager (see our discussion in \citet{grupe2015}, where we also used BLR line width). The large physical size of IC 3599 as shown in Figure\,\ref{swift_image} may support a larger black hole mass as well, but the previous SMBH mass estimates using the various scaling relations are the most restrictive.

\subsection{Conclusions}
Despite the speculation in 1995 by \citet{brandt1995} and \citet{grupe1995a}, and the repeated-TDE model by \citet{campana2015} that the X-ray outburst seen during the RASS could have been caused by a TDE, our long-term \swift\ monitoring campaign suggests otherwise, since the  third outburst predicted by the repeat-TDE model 
in 2019/2020 did not happen. 
The X-ray outbursts seen in 1990 by ROSAT and 2010 by \swift\ are most likely caused by an accretion disk instability as favored by \citet{grupe2015}, since IC 3599 is not a quiescent galaxy but a bona fide AGN. In fact, IC 3599 is an extreme case of an optical changing-look AGN \citep{komossa2023, xu2024}.

Independent of the nature of the big outbursts or their absence in 2019/2020 in particular, we have also studied the long-term X-ray flux and spectral variability properties of IC 3599 during its more quiescent levels. IC 3599 turns out to be highly variable in the X-ray regime. One event stood out:  
We found that IC 3599 exhibited an increase in its X-ray flux by a factor of about 10 in December 2022. This mini flare was followed by a remarkable hardening of the X-ray spectrum in the following months. One possible explanation for such a behavior could be the temporary formation of an accretion disk corona. 

We will continue to monitor IC 3599 with \swift. 
 It depends on the model, if and when a new outburst is expected. For instance, if the previous two outbursts were powered by a disk instability, it would depend on the evolving and variable AGN disk properties, if and when a new instability occurs (note that the X-ray emission of IC 3599 in low-state is not constant but still varies, as in the majority of AGN). If, instead, there was an underlying  strictly periodic process behind the outbursts, like variants of binary SMBH models where for instance a secondary SMBH interacts with the primary's disk once during its orbit, or binary SMBH models where stream-feeding from a circum-binary disk occurs, then the next outburst is expected around the year 2030.   
In case IC 3599 will exhibit another X-ray outburst again, we have a pre-approved ToO program with XMM that would allow us to study the inner-most environment of the central black hole. 

\begin{acknowledgments} 
We would like to thank  \swift\ PI Brad Cenko for approving our continued requests to observe IC 3599 and the \swift\ Science Operations team for executing the observations. 
We would also like to thank the anonymous referee for
useful comments and suggestions. 
This research has made use of the NASA/IPAC 
Extragalactic Database (NED) which is operated by the Jet Propulsion Laboratory,
Caltech, under contract with the National Aeronautics and Space
Administration. 
This work made use of data supplied by the UK Swift Science Data Centre at the University of Leicester \citep{evans2007}. 
This research has made use of the
  XRT Data Analysis Software (XRTDAS) developed under the responsibility
  of the ASI Science Data Center (ASDC), Italy.
This research has made use of data obtained through the High Energy 
Astrophysics Science Archive Research Center Online Service, provided by the 
NASA/Goddard Space Flight Center. 
\end{acknowledgments} 

\vspace{5mm}
\facilities{NASA Neil Gehrels Swift observatory (XRT and UVOT).}

\software{HEASoft (\url{https://heasarc.gsfc.nasa.gov/docs/software/heasoft/}) with XSPEC \citep{arnaud1996}, ESO-MIDAS (\url{https://www.eso.org/sci/software/esomidas/}), 
the R programming language (\url{https://www.r-project.org/}),
and SuperMongo  (\url{https://www.astro.princeton.edu/~rhl/sm/}). 
}

\section{Data Availability}
The raw Swift data of our project are available in the Swift archive at \url{https://swift.gsfc.nasa.gov/archive/}. The reduced results are available at 
 Zenodo: \dataset[10.5281/zenodo.10899673]{\doi{10.5281/zenodo.10899673}}.

\bibliography{ic3599_apj_2024_astroph}

\appendix

\section{Swift Observations}

\begin{table*}
 \caption{\swift\ XRT and UVOT observations  of IC 3599 \label{swift_obs}. 
The full machine-readable table is available on Zenodo: \dataset[10.5281/zenodo.10899673]{\doi{10.5281/zenodo.10899673.}}}
\centering
  \begin{tabular}{cccccrrrrrrr}
  \hline
  \hline
Target ID & Segment &  MJD$^1$ & $T_{\rm start}$ (UT) & $T_{\rm end}$ (UT) & $T_{\rm XRT}$ 
& $T_{\rm  V}$ 
& $T_{\rm  B}$ 
& $T_{\rm  U}$ 
& $T_{\rm  W1}$ 
& $T_{\rm  M2}$ 
& $T_{\rm  W1}$  \\
\hline  
37507 & 001 & 55252.365 & 2010-02-25 07:46 & 2010-02-25 09:44 & 2120 & 176 & 176 & 176 & 352 & 486 & 704 \\
          & 003 & 55333.049 & 2010-05-17 00:58 & 2010-05-17 01:19 & 1234 & 105 & 105 & 105 & 210 & 264 & 421 \\
          & 004 & 56595.160 & 2013-10-30 00:34 & 2013-10-30 07:13 & 3771 & 312 & 312 & 312 & 624 & 795 & 1250 \\
          & 005 & 56602.208 & 2013-11-06 02:24 & 2013-11-06 07:27 & 4832 & --- & --- & --- & --- & 4780 & --- \\
          & 006 & 56742.583 & 2014-03-26 07:50 & 2014-03-26 19:08 & 3596 & 101 & 101 & 101 & 204 & 2630 & 407 \\
37569 & 001 & 56877.306 & 2014-08-08 07:07 & 2014-08-08 07:35 & 1613 & 132 & 132 & 132 & 263 & 381 & 526  \\
          & 002 & 56976.313 & 2014-11-15 04:12 & 2014-11-15 10:49 & 4615 & 107 & 107 & 3390 & 214 & 290 & 430  \\
          & 003 & 56984.375 & 2014-11-23 08:57 & 2014-11-23 09:03 &  352  & --- & --- & 354 & --- & --- & ---  \\ 
          & 004 & 57110.970 & 2015-03-29 21:37 & 2015-03-30 00:50 & 2829 & 229 & 229 & 229 & 460 & 675 & 920  \\
          & 005 & 57209.167 & 2015-07-06 00:33 & 2015-07-06 04:08 & 4840 & 134 & 134 & 134 & 268 & 352 & 3740 \\
          & 006 & 57215.146 & 2015-07-12 01:43 & 2015-07-12 06:40 & 1648 & ---  & 324 & 348 & 699 & --- & 204 \\
          & 007 & 57218.438 & 2015-07-15 07:57 & 2015-07-15 12:52 & 2377 & 193 & 193 & 193 & 387 & 528 & 779  \\
          & 008 & 57220.604 & 2015-07-19 14:18 & 2015-07-19 14:44 & 1528 & 121 & 120 & 121 & 242 & 418 & 484 \\
          & 009 & 57229.149 & 2015-07-26 02:37 & 2015-07-26 04:35 & 1913 & 158 & 158 & 158 & 317 & 430 & 635 \\
          & 010 & 57236.194 & 2015-08-02 03:51 & 2015-08-02 05:43 & 1191 & 112 & 112 & 112 & 226 & 123 & 451  \\
          & 011 & 57239.361 & 2015-08-05 08:34 & 2015-08-05 08:51 &  989  &   81 &  81 &   81 & 160 & 240 & 322  \\
          & 012 & 57243.316 & 2015-08-09 05:01 & 2015-08-09 10:11 & 2000 & 164 & 164 & 164 & 328 & 471 & 658 \\ 
          & 013 & 57327.104 & 2015-11-01 00:03 & 2015-11-01 05:20 & 4697 & 130 & 132 & 132 & 3370 & 379 & 527 \\
          & 014 & 57387.375 & 2015-12-31 04:27 & 2015-12-31 13:43 & 3385 & --- & --- & --- & 1915 & --- & -- \\
          & 015 & 57466.333 & 2016-03-19 00:44 & 2016-03-19 15:11 & 4014 & 105 & 105 & 105 & 211 & 3010 & 423   \\
          & 016 & 57589.506 & 2016-07-20 11:54 & 2016-07-20 12:22 & 1663 & 136 & 136 & 136 & 272 & 409 & 545 \\
          & 017 & 57590.399 & 2016-07-21 07:10 & 2016-07-21 12:02 & 3274 & 264 & 264 & 264 & 530 & 783 & 1058 \\
          & 018 & 57762.924 & 2017-01-09 20:44 & 2016-01-09 23:42 & 2412 & 135 & 135 & 135 & 270 & 1142 & 539 \\
          & 019 & 57765.253 & 2017-01-12 03:03 & 2017-01-12 06:10 & 2515 &   75 &  75 &   75 & 149 & 236 & 1873  \\
          & 020 & 57798.750 & 2017-02-14 14:34 & 2017-02-14 21:11 & 1543 & 117 & 117 & 117 & 234 & 441 & 469  \\
          & 021 & 57810.667 & 2017-02-26 10:34 & 2017-02-26 23:30 & 1471 & 117 & 117 & 117 & 231 & 352 & 463  \\
          & 022 & 57826.507 & 2017-03-12 12:13 & 2017-03-12 12:19 & 3446 & --- & --- &  81 & 259 & --- & --- \\
          & 023 & 57828.439 & 2017-03-16 08:53 & 2017-03-16 12:11 & 1735 & 140 & 140 & 140 & 280 & 414 & 561  \\
          & 024 & 57838.632 & 2017-03-26 14:11 & 2017-03-26 16:05 & 2008 & 165 & 165 & 165 & 329 & 459 & 661 \\
          & 025 & 57852.750 & 2017-04-09 03:30 & 2017-04-09 08:25 & 1331 & 105 & 106 & 106 & 212 & 304 & 425 \\
          & 026 & 57866.910 & 2017-04-23 21:37 & 2017-04-23 22:05 & 1623 & 134 & 134 & 134 & 268 & 390 & 537 \\
          & 027 & 57880.552 & 2017-05-07 06:05 & 2017-05-07 20:25 & 1918 & 155 & 155 & 155 & 310 & 462 & 622 \\
          & 028 & 57894.899 & 2017-05-21 04:42 & 2017-05-22 13:12 & 2093 & 169 & 169 & 169 & 338 & 515 & 677 \\
          & 029 & 57908.858 & 2017-06-04 19:50 & 2017-06-04 21:32 & 1216 & 107 & 107 & 145 & 313 &  78 & 429  \\
          & 030 & 57969.344 & 2017-08-04 00:09 & 2017-08-04 16:16 & 2115 & 171 & 171 & 171 & 273 & 482 & 547 \\   
          & 031 & 58065.042 & 2017-11-08 00:03:& 2017-11-08 01:42 & 1756 & 150 & 150 & 150 & 300 & 514 & 600 \\
10375 & 001 & 58072.705 & 2017-11-15 15:17 & 2017-11-15 18:35 & 3718 & 249 & 381 & 381 & 763 & 689 & 1144 \\  
          & 002 & 58076.618 & 2017-11-19 14:40 & 2017-11-19 14:59 & 1139 &  96 &   96 &  96  & 193 & 233 & 386  \\
          & 003 & 58103.928 & 2017-12-16 22:12 & 2017-12-16 22:24 &  694  &  60 &   60 &  60  & 121 & 172 & 242 \\
          & 004 & 58105.486 & 2017-12-17 23:38 & 2017-12-18 23:41 & 4078 & 340 & 340 & 340 & 679 & 922 & 1361 \\
          & 005 & 58133.514 & 2018-01-15 05:01 & 2018-01-15 19:37 & 4545 & 347 & 347 & 410 & 832 & 1056 & 1390 \\
          & 006 & 58164.500 & 2018-02-15 02:23 & 2018-02-15 21:45 & 4825 & 388 & 388 & 388 & 776 & 1131 & 1554  \\
          & 007 & 58192.318 & 2018-03-15 01:09 & 2018-03-15 14:04 & 4730 & 390 & 387 & 390 & 785 & 983 & 1567 \\
          & 008 & 58223.566 & 2018-04-15 09:35 & 2018-04-15 17:44 & 4877 & 405 & 405 & 405 & 811 & 1042 & 1625 \\
          & 009 & 58253.625 & 2018-05-15 13:10 & 2018-05-15 16:43 & 2944 & 233 & 233 & 233 & 469 & 731 & 935 \\  
          & 010 & 58256.708 & 2018-05-18 16:07 & 2018-05-18 17:57 & 1830 &   89 & 170 & 170 & 341 & 263 & 681 \\
          & 011 & 58284.448 & 2018-06-15 08:57 & 2018-06-15 12:29 & 4935 & 481 & 407 & 407 & 813 & 1134 & 1628 \\
          & 012 & 58315.421 & 2018-07-15 22:05 & 2018-07-16 22:10 & 4635 & 362 & 362 & 362 & 727 & 1139 & 1455 \\
          \hline
\hline
\end{tabular}

$^1$ The Modified Julian Date (MJD) is given for the middle of the observation period. 

\end{table*}

\begin{table*}
\setcounter{table}{0}
\caption{Continued}
\centering
  \begin{tabular}{cccccrrrrrrr}
  \hline
  \hline
Target ID & Segment &  MJD$^1$ & $T_{\rm start}$ (UT) & $T_{\rm end}$ (UT) & $T_{\rm XRT}$ 
& $T_{\rm  V}$ 
& $T_{\rm  B}$ 
& $T_{\rm  U}$ 
& $T_{\rm  W1}$ 
& $T_{\rm  M2}$ 
& $T_{\rm  W1}$  \\
\hline  
10375      & 013 & 58339.170 & 2018-08-09 02:18 & 2018-08-09 05:52 &               4580 & 235 & 235 & 235 & 469 & 749 & 941 \\ 
          & 014 & 58422.882 & 2018-10-31 18:52 & 2018-10-31 23:30 & 4949 & 397 & 397 & 397 & 794 & 1205 & 1592 \\
          & 015 & 58423.389 & 2018-11-01 05:48 & 2018-11-01 12:42 & 5002 & 413 & 413 & 413 & 828 & 1107 & 1659 \\
          & 016 & 58452.387 & 2018-11-30 09:22 & 2018-11-30 14:31 & 4540 & 350 & 317 & 362 & 724 & 1122 & 1450 \\
          & 017 & 58453.542 & 2018-12-01 15:35 & 2018-12-01 17:31 & 2532 & 207 & 207 & 207 & 414 &  584 & 829 \\
          & 018 & 58482.979 & 2018-12-30 22:31 & 2018-12-31 00:34 & 3219 & 268 & 268 & 268 & 536 & 772 & 1074 \\
          & 019 & 58488.059 & 2019-01-05 01:14 & 2019-01-05 01:34 & 1209 & 101 & 101 & 101 & 203 & 258 & 405 \\
          & 020 & 58492.747 & 2019-01-09 17:04 & 2019-01-09 18:47 &  862 & --- & 180 & 180 & 361 & --- & 98  \\
          & 021 & 58496.694 & 2019-01-13 16:27 & 2019-01-13 16:51 & 1409 & 115 & 115 & 115 & 230 & 331 & 460  \\
          & 022 & 58500.344 & 2019-01-17 01:46 & 2019-01-17 15:02 & 4755 & 386 & 386 & 386 & 775 & 1059 & 1552 \\
          & 023 & 58514.403 & 2019-01-31 00:42 & 2019-01-31 20:01 & 1306 & 82 & 141 & 141 & 331 & 219 & 331 \\
& 024 & 58519.486 & 2019-02-05 03:41 & 2019-02-05 19:33 & 1928 & 146 & 146 & 146 & 393 & 439 & 586 \\
          & 025 & 58521.085 & 2019-02-08 01:33 & 2019-02-08 03:22 & 1726 &  140 & 140 & 140 & 281 & 392 & 563 \\
          & 026 & 58531.542 & 2019-02-17 06:50 & 2019-02-17 15:06 & 4448 & 362 & 362 & 362 & 727 & 999 & 1457 \\
          & 027 & 58546.101 & 2019-03-03 13:25 & 2019-03-04 15:03 & 4902 & 398 & 398 & 398 & 802 & 1169 & 1604 \\
          & 028 & 58559.278 & 2019-03-17 00:54 & 2019-03-17 12:19 & 3652 & 263 & 263 & 263 & 531 & 765 & 1059 \\
          & 029 & 58573.531 & 2019-03-31 09:13 & 2019-03-31 16:07 & 4817 & 396 & 396 & 396 & 795 & 1056 & 1590 \\
          & 030 & 58590.403 & 2019-04-17 04:26 & 2019-04-17 14:20 & 4880 & 391 & 391 & 390 & 786 & 1190 & 1575 \\ 
          & 031 & 58620.403 & 2019-05-17 06:22 & 2019-05-17 13:00 & 3985 & 323 & 323 & 323 & 645 &  931 & 1293 \\
          & 032 & 58634.524 & 2019-05-31 02:07 & 2019-05-31 23:16 & 2869 & 164 & 280 & 280 & 684 & 392 & 921 \\
          & 033 & 58651.483 & 2019-06-17 03:25 & 2019-06-17 07:15 & 4905 & 403 & 403 & 403 & 803 & 1149 & 1610\\
          & 034 & 58665.295 & 2019-07-01 02:05 & 2019-07-01 12:04 & 4001 & 348 & 348 & 348 & 696 & 663 & 1397 \\
          & 035 & 58681.139 & 2019-07-17 00:46 & 2019-07-17 06:00 & 4470 & 369 & 369 & 369 & 739 & 1015 & 1479 \\
          & 036 & 58704.569 & 2019-08-09 01:34 & 2019-08-10 01:53 & 3861 & 225 & 225 & 225 & 500 & 736 & 906 \\
          & 037 & 58788.138 & 2019-11-01 01:32 & 2019-11-01 05:07 & 4253 & 345 & 345 & 345 & 690 & 1024 & 1381 \\
          & 038 & 58808.292 & 2019-11-21 04:22 & 2019-11-21 09:35 & 3047 & 250 & 250 & 250 & 499 & 714 & 999 \\ 
          & 040 & 58814.549 & 2019-11-27 00:37 & 2019-11-28 02:23 & 4192 & 346 & 346 & 346 & 693 & 943 & 1387 \\
          & 041 & 58834.819 & 2019-12-17 16:26 & 2019-12-17 22:58 & 3696 & 181 & 376 & 376 & 751 & 612 & 1372 \\
          & 042 & 58865.354 & 2020-01-17 00:33 & 2019-01-17 16:40 & 2642 & 237 & 237 & 237 & 474 & 413 & 947 \\
          & 044 & 58886.552 & 2020-02-07 06:49 & 2019-02-07 19:45 & 2113 & 171 & 172 & 172 & 343 & 465 & 689 \\
          & 045 & 58909.708 & 2020-03-01 13:39 & 2019-03-01 20:23 & 2315 & 188 & 188 & 188 & 376 & 548 & 751 \\
          & 046 & 58943.500 & 2020-04-03 23:27 & 2020-04-04 23:34 & 2233 & 182 & 182 & 182 & 365 & 514 & 731 \\
          & 047 & 58970.292 & 2020-05-01 01:19 & 2020-05-01 12:39 & 2068 & 166 & 166 & 166 & 333 & 474 & 668 \\
          & 048 & 59001.656 & 2020-06-01 15:51 & 2020-06-01 16:02 &  642  &  53 &    54 &  54 & 107 & 125 & 214  \\
          & 049 & 59002.426 & 2020-06-02 06:10 & 2020-06-02 14:19 & 1129 &  87 &    87 &  87 & 178 & 280 & 349 \\
          & 050 & 59031.104 & 2020-07-01 00:07 & 2020-07-01 05:06 & 1743 & 144 & 144 & 144 & 288 & 377 & 575  \\
          & 051 & 59062.650 & 2020-08-01 14:37 & 2020-08-01 16:37 & 2290 & 187 & 187 & 187 & 374 & 534 & 747  \\
          & 052 & 59154.500 & 2020-11-01 03:58 & 2020-11-01 20:03 & 1935 & 158 & 158 & 158 & 315 & 439 & 632 \\
          & 053 & 59184.813 & 2020-12-01 17:02 & 2020-12-01 22:02 & 1626 & 134 & 134 & 134 & 306 & 344 & 534 \\
          & 054 & 59188.892 & 2020-12-05 21:13 & 2020-12-05 21:35 & 1291 & 107 & 107 & 107 & 214 & 286 & 428 \\
          & 055 & 59215.674 & 2021-01-01 15:10 & 2021-01-01 17:11 & 3042 & 246 & 246 & 246 & 493 & 739 & 985 \\
          & 056 & 59246.594 & 2021-02-01 02:47 & 2021-02-01 23:47 & 3112 & 245 & 245 & 245 & 491 & 788 & 985 \\
          & 057 & 59274.087 & 2021-02-28 22:01 & 2021-03-01 06:11 & 1973 & 159 & 159 & 159 & 317 & 469 & 637 \\
          & 058 & 59305.792 & 2021-04-01 18:45 & 2021-04-01 20:37 & 1201 &  98 &   98 &   98 & 195 & 256 & 393 \\
          & 059 & 59340.125 & 2021-05-06 02:18 & 2021-05-06 04:10 & 1538 & 122 & 122 & 122 & 243 & 381 & 485 \\
          & 061 & 59372.774 & 2021-06-06 05:26 & 2021-06-07 21:42 & 1713 & 132 & 132 & 132 & 265 & 425 & 532 \\
          & 062 & 59374.155 & 2021-06-09 03:38 & 2021-06-09 03:48 &  567 &   45 &  45  &  45  &  89  & 131 & 179 \\
          & 063 & 59378.597 & 2021-06-13 14:15 & 2021-06-13 14:28 &  789 &   62 &  62  &  62  & 125 & 197 & 249 \\
\hline
\hline
\end{tabular}

$^1$ The Modified Julian Date (MJD) is given for the middle of the observation period. 

\end{table*}

\begin{table*}
\setcounter{table}{0}
\caption{Continued}
\centering
  \begin{tabular}{cccccrrrrrrr}
  \hline
  \hline
Target ID & Segment &  MJD$^1$ & $T_{\rm start}$ (UT) & $T_{\rm end}$ (UT) & $T_{\rm XRT}$ 
& $T_{\rm  V}$ 
& $T_{\rm  B}$ 
& $T_{\rm  U}$ 
& $T_{\rm  W1}$ 
& $T_{\rm  M2}$ 
& $T_{\rm  W1}$  \\
\hline  
10375     & 064 & 59396.359 & 2021-07-01 06:07 & 2021-07-01 11:08 &               1508 & 118 & 118 & 118 & 237 & 375 & 474 \\
          & 065 & 59427.389 & 2021-08-01 05:56 & 2021-08-01 12:38 & 2290 & 182 & 182 & 182 & 362 & 580 & 725 \\
          & 066 & 59522.399 & 2021-11-04 06:55 & 2021-11-04 12:10 & 3366 & 273 & 273 & 273 & 548 & 812 & 1096 \\
          & 067 & 59527.740 & 2021-11-09 12:47 & 2021-11-09 22:45 & 2277 & 180 & 180 & 180 & 360 & 572 & 721 \\
          & 068 & 59552.469 & 2021-12-04 05:47 & 2021-12-04 18:37 & 2605 & 178 & 247 & 284 & 568 & 526 & 710 \\
          & 069 & 59556.222 & 2021-12-08 05:19 & 2021-12-08 05:23 &  245  & ---  & ---  & 34    & 203 & --- & --- \\
          & 070 & 59560.128 & 2021-12-12 02:54 & 2021-12-12 03:13 & 1126 &   94 &   94 &   94 & 188 & 239 & 377 \\
          & 071 & 59583.910 & 2022-01-04 21:02 & 2022-01-04 23:07 & 3479 & 281 & 281 & 281 & 563 & 853 & 1128 \\
          & 073 & 59674.611 & 2022-04-05 13:46 & 2022-04-05 15:34 & 1181 &   91 &  91 &  91 & 183 & 296 & 366 \\
          & 074 & 59703.993 & 2022-05-04 05:44 & 2022-05-04 12:02 & 1888 & 146 & 146 & 146 & 294 & 435 & 588  \\
          & 075 & 59704.432 & 2022-05-05 10:27 & 2022-05-05 16:43 & 1721 & 141 & 141 & 141 & 281 & 380 & 563 \\
          & 076 & 59734.146 & 2022-06-03 17:50 & 2022-06-04 11:25 & 2375 & 197 & 197 & 197 & 396 & 514 & 792 \\
          & 077 & 59764.390 & 2022-07-04 06:08 & 2022-07-04 12:37 & 4163 & 328 & 328 & 328 & 656 & 1063 & 1313  \\
          & 078 & 59795.644 & 2022-08-04 01:49 & 2022-08-04 21:04 & 3334 & 266 & 266 & 266 & 530 & 779 & 1064 \\
          & 079 & 59883.310  & 2022-10-31 07:19 & 2022-10-31 15:47 & 2055 & 175 & 175 & 175 & 348 & 416 & 693 \\
          & 080 &  59914.585  & 2022-12-01 03:09 & 2022-12-01 20:59 & 1024 &  44 &  89  & 181 & 363 & 122 & 179  \\
          & 081 & 59919.955   & 2022-12-06 22:51 & 2022-12-06 22:56 &  405 &  --- &   69 &   69 & 138 & ---  & 109   \\
          & 082 & 59923.339   & 2022-12-10 08:05 & 2022-12-10 08:12 &  414 &  37 &   37 &   37 &  75  & 49  & 150   \\
          & 083 & 59936.658   & 2022-12-23 15:34 & 2022-12-23 16:03 & 1716 & 141 & 141 & 141 & 280 & 404 & 562   \\
          & 084 & 59944.535   & 2022-12-31 12:37 & 2022-12-31 13:04 & 1576 & 135 & 135 & 135 & 270 & 318 & 539   \\
          & 085 & 59948.812   & 2023-01-04 18:54 & 2023-01-04 20:05 &  817 & --- & 157 & 157 & 314 & --- & 145     \\
          & 086 & 59949.630   & 2023-01-05 15:00 & 2023-01-05 15:16 &  913 &  79 &  79 &  79 & 157 & 192 & 315    \\
          & 087 & 59950.838   & 2023-01-06 19:59 & 2023-01-06 20:15 &  926 & 76  &  76 &  76 & 151 & 211 & 303     \\
          & 088 & 59951.623   & 2023-01-07 14:49 & 2023-01-07 15:05 &  844 & 71 &  71 &  71 &  142 & 172 & 283     \\
          & 089 & 59952.411   & 2023-01-08 09:51 & 2023-01-08 10:05 &  804 & 65 &  65 &  65 &  130 & 187 & 261     \\
          & 090 & 59953.011   & 2023-01-09 00:09 & 2023-01-09 00:23 &  794 & 70 &  70 &  70 & 139 & 134 & 278      \\
          & 091 & 59954.007   & 2023-01-10 00:05 & 2023-01-10 00:17 &  694 & 71 &  71 &  71 & 142 &  30 & 283      \\
          & 092 & 59955.596   & 2023-01-11 14:11 & 2023-01-11 14:25 &  819 & 69 &  69 &  69 & 137 & 163 & 276      \\
          & 093 & 59956.868   & 2023-01-12 20:42 & 2023-01-12 20:58 &  914 & 75 &  75 &  75 & 149 & 202 & 300     \\
          & 094 & 59957.791   & 2023-01-13 18:52 & 2023-01-13 19:06 &  824 & 67 &  67 &  67 & 134 & 188 & 269      \\
          & 095 & 59964.619   & 2023-01-20 06:52 & 2023-01-20 22:51 & 1156 & 97 & 97 &  97 & 194 & 221 & 389      \\
          & 096 & 59974.341   & 2023-01-30 08:05 & 2023-01-30 08:17 &  660 & 68 & 68 &  68  & 136 & --- & 272     \\
          & 097 & 59975.276   & 2023-01-31 06:30 & 2023-01-31 09:57 & 1503 & 46 &  170 & 170 & 341 & 156 & 564     \\
          & 098 & 59977.793   & 2023-02-02 18:59 & 2023-02-02 19:08 &  534 & 47 & 47 & 47 & 93 & 81 & 188    \\
          & 099 & 59984.472   & 2023-02-09 11:14 & 2023-02-09 11:25 &  642 & 50 & 50 & 50 & 100 & 160 & 201     \\
          & 100 & 59994.269   & 2023-02-19 06:21 & 2023-02-19 06:34 &  789 & 68 & 68 & 68 & 136 & 143 & 272 \\
          & 101 & 60004.532   & 2023-03-01 12:41 & 2023-03-01 12:51 &  604 & 54 & 54 & 54 & 108 & 87  & 216 \\
          & 102 & 60009.765   & 2023-03-06 18:14 & 2023-03-06 18:31 &  979 & --- & 128 & 128 & 257 & --- & 439 \\
          & 103 & 60014.532   & 2023-03-11 07:54 & 2023-03-11 17:38 & 1174 & 101 & 101 & 101 & 201 & 202 & 403 \\
          & 104 & 60024.925   & 2023-03-21 22:05 & 2023-03-21 22:19 &  821 & 66 & 66 & 66 & 132 & 195 & 264 \\
          & 105 & 60034.588   & 2023-03-31 07:46 & 2023-03-31 20:29 & 1400 & 122 & 122 & 122 & 244 & 238 & 489 \\
          & 106 & 60044.968   & 2023-04-10 23:05 & 2023-04-10 23:24 & 1096 & 90 & 90 & 90 & 182 & 240 & 364 \\
          & 107 & 60054.563   & 2023-04-20 13:24 & 2023-04-20 13:38 &  817 & 68 & 68 & 68 & 137 & 168 & 273 \\
          & 108 & 60065.357   & 2023-05-01 08:21 & 2023-05-01 08:48 & 1613 & 132 & 132 & 132 & 264 & 384 & 527 \\
          & 109 & 60095.205   & 2023-05-31 04:05 & 2023-05-31 05:46 & 1214 & 159 &  65 &  65 & 318 & 408 & 637 \\
          & 110 & 60126.288   & 2023-07-01 02:52 & 2023-07-01 10:58 & 1900 & 165 & 165 & 165 & 328 & 348 & 657 \\
          & 111 & 60156.127   & 2023-07-31 02:50 & 2023-07-31 03:16 & 1518 & 124 & 124 & 124 & 249 & 358 & 498 \\
\hline
\hline
\end{tabular}

$^1$ The Modified Julian Date (MJD) is given for the middle of the observation period. 

\end{table*}

\begin{table*}
\caption{Merged XRT data as shown in Figure \ref{swift_xrt_lc} \label{xrt_merge}}
\begin{tabular}{lcrcc}
\hline
\hline
MJD Range & MJD$_{\rm center}$ &  bin width$^1$ & $F_{0.3-10 keV}^2$ & HR$^3$ \\
\hline
56595 - 57243 & 56916.0 & 648.0 & 0.943\plm0.120 & +0.00\plm0.10 \\
57327 - 57969 & 57648.0 & 642.0 & 0.906\plm0.100 & -0.36\plm0.15  \\
58065 - 58339 & 58202.0 & 137.0 & 0.615\plm0.059 & -0.22\plm0.12  \\
58422 - 58704 & 58563.0 & 141.0 & 1.008\plm0.065 & -0.39\plm0.12  \\
58788 - 59062 & 58925.0 & 137.0 & 1.870\plm0.131 & -0.66\plm0.07  \\
59154 - 59427 & 59290.5 & 136.5 & 2.170\plm0.219 & -0.54\plm0.07  \\
59527 - 59795 & 59661.0 & 134.0 & 3.080\plm0.260 & -0.36\plm0.08  \\
59883 - 59975$^4$ & 59929.0 & 46.0  & 4.557\plm0.817 & -0.68\plm0.05 \\  
59977 - 60065$^5$ & 60021.0 & 44.0  & 1.340\plm0.558 & +0.22\plm0.18  \\

\hline
\hline
\end{tabular}

$^{1}$ given in days \\

$^{2}$ The observed 0.3-10 keV fluxes are given in units of $10^{-16}$ W m$^{-2}$. \\

$^{3}$ The hardness ratio is defined as $HR = \frac{hard - soft}{hard+soft}$ with the soft and hard bands in the 0.3-1.0 and 1.0-10 keV bands, respectively, applying the BEHR program by citet{park2006}

$^{4}$ The data from the mini flare in December 2022.

$^{5}$ After flare data from February to May 2023

\end{table*}

\begin{table*}
 \centering
 \caption{\swift\ XRT and UVOT fluxes of IC 3599 \label{swift_res}
The full machine-readable table is available on Zenodo: \dataset[10.5281/zenodo.10899673]{\doi{10.5281/zenodo.10899673}}.
 }
  \begin{tabular}{crcrrrrrr}
  \hline
  \hline

  MJD & $F_{\rm 0.3-10 keV}^{1}$  & HR$^{2}$ 
& $F_{\rm V }^{3}$ 
& $F_{\rm B }^{3}$ 
& $F_{\rm U }^{3}$ 
& $F_{\rm W1 }^{3}$ 
& $F_{\rm M2 }^{3}$ 
& $F_{\rm W2 }^{3}$  \\
\hline  
 55252.365 & 105.8\plm6.00  & -0.91\plm0.03 & 13.03\plm0.48 & 10.01\plm0.37 & 7.25\plm0.27 & 5.83\plm0.21 & 5.69\plm0.26 & 5.13\plm0.19   \\
 55333.049 & 29.09\plm3.14 & -0.86\plm0.07 & 11.77\plm0.55 & 9.47\plm0.35  & 6.80\plm0.25 & 5.22\plm0.24 & 5.48\plm0.31 & 5.04\plm0.23	 \\
 56595.160 & 1.15\plm0.45$^4$  & -0.18\plm0.30 &  8.85\plm0.33 & 7.31\plm0.27  & 3.60\plm0.16 & 2.71\plm0.12 & 2.24\plm0.12 & 2.39\plm0.08	 \\
 56602.208 & 1.08\plm0.40$^4$  & 	 ---	   &	   ---     &	 ---	   &	  ---	  &	 ---	 & 2.50\plm0.09 &      --- 	 \\
 56742.583 & 0.73\plm0.33$^4$  & 	 ---   	   &  9.52\plm0.53 & 7.45\plm0.27  & 3.91\plm0.18 & 2.66\plm0.17 & 2.48\plm0.09 & 2.45\plm0.13	 \\
 56877.306 &         ---         & 	 ---   	   &  9.35\plm0.43 & 7.18\plm0.26  & 3.77\plm0.17 & 2.79\plm0.15 & 2.62\plm0.17 & 2.52\plm0.11	 \\
 56976.188 & 1.51\plm0.40$^4$  & 	 ---   	   &  9.10\plm0.51 & 7.66\plm0.35  & 3.56\plm0.13 & 2.38\plm0.15 & 2.02\plm0.15 & 2.28\plm0.12	 \\
 57110.970 & 0.47\plm0.22$^4$  & 	 ---   	   &  8.77\plm0.32 & 6.92\plm0.25  & 3.43\plm0.16 & 2.36\plm0.13 & 2.08\plm0.11 & 2.08\plm0.09	 \\
 57209.167 &         ---         & 	 ---   	   &  9.35\plm0.43 & 7.52\plm0.28  & 3.53\plm0.16 & 2.89\plm0.16 & 2.74\plm0.18 & 2.47\plm0.09	 \\
 57215.146 & 1.40\plm0.75$^4$  & 	 ---   	   &	   ---     & 6.98\plm0.26  & 3.53\plm0.13 & 2.52\plm0.11 &      ---	& 2.32\plm0.15	 \\
 57218.438 & 1.00\plm0.55$^4$  & 	 ---   	   &  9.02\plm0.42 & 7.18\plm0.26  & 3.34\plm0.15 & 2.66\plm0.15 & 2.32\plm0.15 & 2.36\plm0.11	 \\
 57220.604 & 1.55\plm0.75$^4$  & 	 ---   	   &  9.53\plm0.44 & 7.18\plm0.26  & 3.73\plm0.17 & 2.52\plm0.16 & 2.53\plm0.16 & 2.59\plm0.14	 \\
 57229.149 & 1.15\plm0.75$^4$  & 	 ---   	   &  8.53\plm0.39 & 7.38\plm0.34  & 3.70\plm0.17 & 2.69\plm0.15 & 2.80\plm0.18 & 2.54\plm0.11	 \\
 57236.194 & 1.55\plm1.12$^4$  & 	 ---   	   &  9.71\plm0.54 & 7.80\plm0.36  & 3.63\plm0.20 & 2.61\plm0.17 & 2.74\plm0.26 & 2.30\plm0.13	 \\
 57239.361 & 2.90\plm1.20$^4$  & 	 ---   	   &  9.80\plm0.55 & 7.80\plm0.36  & 3.34\plm0.22 & 3.03\plm0.20 & 2.43\plm0.18 & 2.52\plm0.14	 \\
 57243.316 &      ---            & 	 ---   	   &  9.80\plm0.45 & 7.59\plm0.28  & 3.77\plm0.17 & 2.74\plm0.15 & 2.53\plm0.16 & 2.59\plm0.12	 \\
 57327.104 & 0.86\plm0.20$^4$  & 	 ---   	   &  9.35\plm0.52 & 7.25\plm0.27  & 3.53\plm0.19 & 2.59\plm0.09 & 2.72\plm0.18 & 2.45\plm0.13	 \\
 57387.375 & 1.35\plm0.45$^4$  & 0.01\plm0.37  &       ---	   &      ---	   &      ---     & 2.71\plm0.10 &      ---             ---	 \\
 57466.333 & 0.68\plm0.30$^4$  &     ---       &  8.53\plm0.48 & 6.67\plm0.31  & 3.43\plm0.16 & 2.47\plm0.16 & 2.46\plm0.06 & 2.32\plm0.13	 \\
 57590.500 & 1.00\plm0.35$^4$  & -0.11\plm0.35 &  9.27\plm0.34 & 7.18\plm0.26  & 3.60\plm0.13 & 2.76\plm0.12 & 2.48\plm0.14 & 2.39\plm0.11	 \\
 57762.924 & 3.76\plm1.01$^4$  & -0.59\plm0.17 &  9.35\plm0.43 & 7.18\plm0.26  & 3.84\plm0.18 & 2.71\plm0.17 & 2.39\plm0.11 & 2.59\plm0.12	 \\
 57765.253 &      ---            & 	 ---	   &  9.80\plm0.55 & 6.79\plm0.31  & 3.80\plm0.21 & 2.74\plm0.18 & 2.50\plm0.19 & 2.15\plm0.80	\\
 57798.750 &      ---            & 	 ---	   &  9.71\plm0.45 & 7.45\plm0.27  & 3.84\plm0.18 & 2.74\plm0.18 & 2.24\plm0.14 & 2.62\plm0.14	\\
 57810.667 & 2.15\plm1.00$^4$  & 	 ---	   & 10.16\plm0.47 & 7.38\plm0.27  & 4.13\plm0.19 & 2.89\plm0.19 & 2.65\plm0.07 & 2.54\plm0.14	\\
 57826.507 & 1.76\plm0.81$^4$  & 	 ---	   &  8.69\plm0.40 & 7.31\plm0.27  & 3.98\plm0.18 & 2.79\plm0.15 & 2.82\plm0.18 & 2.69\plm0.12	\\
 57838.632 &       ---           & 	 ---	   &  8.93\plm0.41 & 7.18\plm0.26  & 3.53\plm0.16 & 2.57\plm0.14 & 2.65\plm0.17 & 2.36\plm0.11	\\
 57852.750 & 1.82\plm1.69$^4$  & 	 ---	   &  9.62\plm0.54 & 7.11\plm0.26  & 3.43\plm0.19 & 2.52\plm0.16 & 2.46\plm0.16 & 2.32\plm0.13	\\
 57866.910 & 1.49\plm1.07$^4$  & 	 ---	   &  9.53\plm0.44 & 7.18\plm0.26  & 3.91\plm0.18 & 2.89\plm0.16 & 2.60\plm0.17 & 2.47\plm0.14	\\
 57880.552 &        ---          & 	 ---	   &  9.71\plm0.45 & 6.92\plm0.25  & 3.40\plm0.15 & 2.89\plm0.16 & 2.22\plm0.14 & 2.32\plm0.10	\\
 57894.899 & 2.14\plm0.81$^4$  & 	 ---	   &  8.61\plm0.40 & 7.18\plm0.26  & 3.77\plm0.17 & 2.74\plm0.15 & 2.50\plm0.16 & 2.52\plm0.11	\\
 57908.858 &       ---           & 	 ---	   &  9.62\plm0.54 & 7.38\plm0.27  & 3.63\plm0.17 & 2.84\plm0.16 & 2.28\plm0.31 & 2.36\plm0.13	\\
 57969.344 & 1.81\plm0.81$^4$  & 	 ---	   &  9.35\plm0.43 & 7.05\plm0.26  & 3.63\plm0.17 & 2.64\plm0.17 &      ---     & 2.57\plm0.12	\\
 58065.042 &         ---         & 	 ---	   &  9.10\plm0.51 & 6.92\plm0.25  & 3.34\plm0.15 & 2.19\plm0.12 & 1.99\plm0.13 & 1.73\plm0.97	\\
 58072.705 & 1.50\plm0.54$^4$  & 	 ---	   &  9.62\plm0.35 & 7.11\plm0.26  & 3.56\plm0.13 & 2.57\plm0.12 & 2.48\plm0.14 & 2.30\plm0.10	\\
 58076.618 &        ---          & 	 ---	   &  9.98\plm0.56 & 6.73\plm0.31  & 3.84\plm0.21 & 2.52\plm0.16 & 2.48\plm0.18 & 2.36\plm0.13	\\
 58103.928 &        ---          & 	 ---   	   & 10.30\plm0.78 & 7.11\plm0.40  & 4.05\plm0.26 & 2.81\plm0.21 & 2.87\plm0.24 & 2.10\plm0.13	 \\
 58105.486 & 1.18\plm0.46$^4$  & 	 ---   	   &  9.01\plm0.33 & 7.18\plm0.26  & 3.60\plm0.13 & 2.50\plm0.11 & 2.55\plm0.11 & 2.45\plm0.09	 \\
 58133.514 &        ---          & 	 ---   	   &  9.27\plm0.34 & 7.11\plm0.26  & 3.56\plm0.13 & 2.57\plm0.12 & 2.50\plm0.11 & 2.47\plm0.09	 \\
 58164.500 & 1.02\plm0.43$^4$  & -0.63\plm0.37 &  9.35\plm0.34 & 7.18\plm0.26  & 3.70\plm0.13 & 2.57\plm0.12 & 2.67\plm0.12 & 2.66\plm0.09	 \\
 58192.318 & 1.74\plm0.46$^4$  & -0.31\plm0.24 &  8.93\plm0.33 & 7.11\plm0.26  & 3.77\plm0.14 & 2.59\plm0.12 & 2.57\plm0.12 & 2.54\plm0.09	 \\
 58223.566 & 1.58\plm0.52$^4$  & -0.57\plm0.32 &  9.27\plm0.34 & 7.11\plm0.26  & 3.80\plm0.14 & 2.61\plm0.12 & 2.48\plm0.11 & 2.43\plm0.09	 \\
 58253.625 & 2.12\plm0.70  & -0.32\plm0.30 &  9.89\plm0.36 & 7.38\plm0.27  & 3.66\plm0.13 & 2.81\plm0.13 & 2.53\plm0.14 & 2.54\plm0.11	 \\
 58256.708 & 1.48$^4$\plm0.90  &       ---     &  9.62\plm0.54 & 7.05\plm0.26  & 3.87\plm0.18 & 2.69\plm0.15 & 2.57\plm0.17 & 2.45\plm0.11	 \\
 58284.448 & 1.05\plm0.58  & +0.03\plm0.26 &  9.10\plm0.33 & 7.38\plm0.27  & 3.66\plm0.13 & 2.79\plm0.13 & 2.55\plm0.11 & 2.54\plm0.09	 \\
 58315.421 & 0.83\plm0.30$^4$  & -0.48\plm0.36 &  9.53\plm0.35 & 7.31\plm0.27  & 3.77\plm0.17 & 2.74\plm0.12 & 2.57\plm0.12 & 2.45\plm0.06	 \\
 \hline
\hline
\end{tabular}

$^{1}$ The observed 0.3-10 keV fluxes are given in units of $10^{-16}$ W m$^{-2}$. \\

$^{2}$ The hardness ratios are as defined in Table \ref{xrt_merge}

$^{3}$ The Fluxes in the UVOT filters are given in units of $10^{-15}$ W m$^{-2}$. \\

$^4$ 0.3-10 keV fluxes estimated from the course rates based on the latest possible spectral analysis. 

\end{table*}

\begin{table*}
\setcounter{table}{2}
\caption{Continued}
 \centering
  \begin{tabular}{crcrrrrrr}
  \hline
  \hline

  MJD & $F_{\rm 0.3-10 keV}^{1}$  & HR$^{2}$ 
& $F_{\rm V }^{3}$ 
& $F_{\rm B }^{3}$ 
& $F_{\rm U }^{3}$ 
& $F_{\rm W1 }^{3}$ 
& $F_{\rm M2 }^{3}$ 
& $F_{\rm W2 }^{3}$  \\
\hline  
 58339.170 & 0.64\plm0.25  & -0.46\plm0.22 & 10.30\plm0.38 & 7.05\plm0.26  & 3.98\plm0.18 & 2.71\plm0.15 & 2.80\plm0.15 & 2.66\plm0.12	 \\
 58422.882 &        ---          &       ---     &  9.44\plm0.35 & 7.11\plm0.26  & 3.80\plm0.17 & 2.69\plm0.12 & 2.67\plm0.12 & 2.54\plm0.09	 \\
 58423.389 & 0.68\plm0.23$^4$  & -0.44\plm0.23 &  9.62\plm0.54 & 7.18\plm0.26  & 3.87\plm0.14 & 2.66\plm0.12 & 2.50\plm0.11 & 2.52\plm0.09	 \\
 58452.387 & 1.36\plm0.45$^4$  & -0.66\plm0.28 &  9.35\plm0.34 & 7.11\plm0.26  & 3.66\plm0.13 & 2.64\plm0.12 & 2.37\plm0.11 & 2.50\plm0.09	 \\
 58453.542 & 1.29\plm0.40$^4$  & -0.30\plm0.19 &  9.89\plm0.36 & 7.25\plm0.27  & 3.53\plm0.13 & 2.61\plm0.14 & 2.48\plm0.14 & 2.47\plm0.11	 \\
 58482.979 & 1.50\plm0.50$^4$  & -0.47\plm0.25 &  9.80\plm0.36 & 7.18\plm0.26  & 3.60\plm0.16 & 2.81\plm0.13 & 2.69\plm0.15 & 2.47\plm0.11	 \\
 58488.059 &       ---           &       ---     & 10.25\plm0.57 & 7.66\plm0.28  & 3.80\plm0.21 & 2.54\plm0.16 & 2.57\plm0.17 & 2.54\plm0.14	 \\
 58492.747 &       ---           &       ---     &	   ---     & 7.11\plm0.26  & 3.77\plm0.17 & 2.74\plm0.15 &      ---	& 2.57\plm0.24	 \\
 58496.694 &       ---           &       ---     & 10.25\plm0.48 & 7.05\plm0.33  & 4.21\plm0.19 & 2.79\plm0.18 & 2.55\plm0.16 & 2.64\plm0.14	 \\
 58500.344 & 0.93\plm0.31$^4$  & -0.75\plm0.15 &  9.44\plm0.35 & 7.18\plm0.26  & 3.84\plm0.14 & 2.59\plm0.12 & 2.57\plm0.12 & 2.45\plm0.09	 \\
 58514.403 & 1.41\plm0.52$^4$  & +0.12\plm0.34 &  9.62\plm0.54 & 7.31\plm0.27  & 3.84\plm0.18 & 2.66\plm0.15 & 2.32\plm0.17 & 2.64\plm0.14	 \\ 58519.486 & 1.41\plm0.35  & -0.96\plm0.04 &  9.27\plm0.43 & 7.31\plm0.27  & 3.87\plm0.18 & 2.69\plm0.15 & 2.48\plm0.16 & 2.69\plm0.12	 \\
 58521.085 &        ---          &       ---     &  9.89\plm0.46 & 7.73\plm0.28  & 3.91\plm0.18 & 2.76\plm0.15 & 2.57\plm0.17 & 2.54\plm0.11	 \\
 58531.542 & 1.08\plm0.26$^4$  & -0.63\plm0.20 &  9.61\plm0.35 & 7.31\plm0.27  & 3.84\plm0.18 & 2.81\plm0.13 & 2.41\plm0.11 & 2.47\plm0.09	 \\
 58546.101 & 1.78\plm0.30  & -0.30\plm0.24 &  9.98\plm0.27 & 7.31\plm0.27  & 3.91\plm0.14 & 2.74\plm0.12 & 2.37\plm0.11 & 2.62\plm0.09	 \\
 58559.278 & 1.45\plm0.50$^4$  & -0.20\plm0.32 &  9.53\plm0.35 & 7.31\plm0.27  & 3.91\plm0.14 & 2.76\plm0.12 & 2.46\plm0.13 & 2.39\plm0.11	 \\
 58573.531 & 1.38\plm0.38$^4$  & -0.15\plm0.28 &  9.35\plm0.34 & 7.18\plm0.26  & 3.56\plm0.13 & 2.66\plm0.12 & 2.48\plm0.11 & 2.43\plm0.09	 \\
 58590.403 & 2.31\plm0.42  & -0.95\plm0.05 &  9.18\plm0.34 & 7.31\plm0.27  & 3.73\plm0.13 & 2.54\plm0.11 & 2.43\plm0.11 & 2.43\plm0.09	 \\
 58620.403 & 2.05\plm0.63$^4$  & -0.45\plm0.27 &  9.53\plm0.35 & 7.31\plm0.27  & 3.77\plm0.17 & 2.61\plm0.12 & 2.55\plm0.11 & 2.39\plm0.08	 \\
 58634.524 & 1.15\plm0.45$^4$  & -0.87\plm0.13 &  9.27\plm0.43 & 7.45\plm0.27  & 3.73\plm0.13 & 2.79\plm0.13 & 2.50\plm0.16 & 2.52\plm0.11	 \\
 58651.483 & 1.81\plm0.50  & -0.07\plm0.21 &  9.62\plm0.35 & 7.38\plm0.27  & 3.84\plm0.14 & 2.66\plm0.12 & 2.62\plm0.12 & 2.50\plm0.09	 \\
 58665.295 & 1.98\plm0.69  & -0.51\plm0.23 &  9.02\plm0.33 & 7.05\plm0.26  & 3.47\plm0.12 & 2.64\plm0.12 & 2.28\plm0.12 & 2.28\plm0.08	 \\
 58681.139 & 0.96\plm0.32$^4$  & +0.30\plm0.35 &  9.27\plm0.34 & 7.38\plm0.27  & 3.60\plm0.13 & 2.64\plm0.12 & 2.28\plm0.10 & 2.28\plm0.08	 \\
 58704.569 & 0.58\plm0.31$^4$  & -0.42\plm0.24 &  9.44\plm0.35 & 7.38\plm0.27  & 4.02\plm0.18 & 2.71\plm0.12 & 2.62\plm0.14 & 2.45\plm0.11	 \\
 58788.138 & 2.04\plm0.45  & -0.91\plm0.09 &  9.71\plm0.36 & 7.45\plm0.27  & 3.70\plm0.17 & 2.71\plm0.12 & 2.60\plm0.12 & 2.41\plm0.90	 \\
 58808.292 & 1.77\plm0.54  & -0.47\plm0.29 &  9.98\plm0.37 & 7.11\plm0.26  & 3.80\plm0.17 & 2.59\plm0.12 & 2.50\plm0.14 & 2.50\plm0.11	 \\
 58814.549 & 3.32\plm0.67  & -0.67\plm0.16 & 10.16\plm0.37 & 7.31\plm0.27  & 3.87\plm0.18 & 2.74\plm0.12 & 2.46\plm0.11 & 2.52\plm0.09	 \\
 58834.819 & 1.04\plm0.42$^4$  & +0.06\plm0.38 &  9.53\plm0.44 & 7.45\plm0.27  & 3.77\plm0.14 & 2.57\plm0.12 & 2.53\plm0.14 & 2.36\plm0.08	 \\
 58865.354 & 1.82\plm0.66$^4$  & -0.71\plm0.29 &  9.62\plm0.35 & 7.38\plm0.27  & 3.70\plm0.13 & 2.57\plm0.14 & 2.41\plm0.13 & 2.26\plm0.10	 \\
 58886.552 & 3.28\plm0.88$^4$  & -0.76\plm0.24 & 10.25\plm0.48 & 7.45\plm0.27  & 3.91\plm0.18 & 2.69\plm0.15 & 2.74\plm0.18 & 2.66\plm0.12	 \\
 58909.708 & 3.16\plm0.88$^4$  & -0.35\plm0.25 &  9.71\plm0.45 & 7.25\plm0.27  & 4.17\plm0.15 & 2.79\plm0.15 & 2.85\plm0.16 & 2.64\plm0.12	 \\
 58943.500 & 1.93\plm0.81$^4$  & -0.95\plm0.05 &  9.35\plm0.43 & 7.59\plm0.28  & 4.21\plm0.15 & 2.84\plm0.16 & 2.50\plm0.14 & 2.47\plm0.11	 \\
 58970.292 & 2.78\plm0.96$^4$  & -0.78\plm0.23 &  9.44\plm0.71 & 8.02\plm0.45  & 4.13\plm0.27 & 2.43\plm0.20 & 2.72\plm0.26 & 2.45\plm0.16	 \\
 59002.426 & 4.75\plm1.73  & -0.79\plm0.21 & 10.54\plm0.59 & 7.45\plm0.34  & 3.98\plm0.22 & 2.76\plm0.18 & 2.46\plm0.16 & 2.45\plm0.16	 \\
 59031.104 & 2.64\plm0.79$^4$  &       ---     &  9.62\plm0.45 & 7.59\plm0.28  & 4.02\plm0.18 & 2.79\plm0.18 & 2.32\plm0.15 & 2.59\plm0.12	 \\
 59062.650 & 1.17\plm0.37$^4$  & -0.72\plm0.28 &  8.30\plm0.46 & 5.50\plm0.25  & 2.51\plm0.21 & 2.38\plm0.18 & 2.55\plm0.16 & 2.39\plm0.11	 \\
 59154.500 & 1.92\plm0.60  & -0.58\plm0.22 &  9.80\plm0.45 & 7.38\plm0.27  & 4.05\plm0.19 & 2.66\plm0.15 & 2.53\plm0.16 & 2.43\plm0.11	 \\
 59184.813 & 0.75\plm0.35$^4$  &       ---     &  9.02\plm0.50 & 6.92\plm0.25  & 3.91\plm0.18 & 2.57\plm0.14 & 2.37\plm0.15 & 2.36\plm0.13	 \\
 59188.892 & 1.95\plm0.48$^4$  &       ---     &  9.80\plm0.55 & 7.80\plm0.29  & 3.80\plm0.21 & 2.66\plm0.17 & 2.65\plm0.17 & 2.28\plm0.12	 \\
 59215.674 & 1.97\plm0.73$^4$  &       ---     &  9.44\plm0.35 & 7.38\plm0.27  & 3.66\plm0.13 & 2.74\plm0.12 & 2.48\plm0.14 & 2.39\plm0.11	 \\
 59246.594 & 1.66\plm0.60  & -0.29\plm0.24 &  9.62\plm0.35 & 7.11\plm0.26  & 3.91\plm0.14 & 2.79\plm0.13 & 2.62\plm0.14 & 2.69\plm0.12	 \\
 59274.087 & 0.86\plm0.39$^4$  &       ---     &  9.44\plm0.44 & 7.38\plm0.27  & 3.60\plm0.16 & 2.38\plm0.15 & 2.26\plm0.15 & 2.28\plm0.10	 \\
 59305.792 & 1.83\plm0.71$^4$  &       ---     &  9.80\plm0.55 & 7.45\plm0.34  & 3.56\plm0.20 & 2.66\plm0.20 & 2.32\plm0.17 & 2.59\plm0.14	 \\
 59340.125 & 1.41\plm0.52$^4$  &       ---     &  9.62\plm0.45 & 7.52\plm0.28  & 3.80\plm0.17 & 2.45\plm0.16 & 2.26\plm0.15 & 2.26\plm0.12	 \\ 
 59372.774 & 1.18\plm0.41$^4$  & -0.20\plm0.30 &  9.27\plm0.43 & 7.38\plm0.27  & 3.73\plm0.17 & 2.81\plm0.18 & 2.65\plm0.22 & 2.54\plm0.14	 \\
 59374.155 &        ---          &       ---     &  9.53\plm0.72 & 7.52\plm0.42  & 4.45\plm0.33 & 2.92\plm0.25 & 2.24\plm0.23 & 2.26\plm0.17	 \\
 
\hline
\hline
\end{tabular}

$^{1}$ The observed 0.3-10 keV fluxes are given in units of $10^{-16}$ W m$^{-2}$. \\

$^{2}$ The hardness ratio is defined as $HR = \frac{hard - soft}{hard+soft}$ with the soft and hard bands in the 0.3-1.0 and 1.0-10 keV bands, respectively, appying the BEHR program by \citet{park2006}

$^{3}$ The Fluxes in the UVOT filters are given in units of $10^{-15}$ W m$^{-2}$. \\

$^4$ 0.3-10 keV fluxes estimated from the course rates based on the latest possible spectral analysis.

\end{table*}

\begin{table*}
\setcounter{table}{2}
\caption{Continued}
 \centering
  \begin{tabular}{crcrrrrrr}
  \hline
  \hline

  MJD & $F_{\rm 0.3-10 keV}^{1}$  & HR$^{2}$ 
& $F_{\rm V }^{3}$ 
& $F_{\rm B }^{3}$ 
& $F_{\rm U }^{3}$ 
& $F_{\rm W1 }^{3}$ 
& $F_{\rm M2 }^{3}$ 
& $F_{\rm W2 }^{3}$  \\
\hline  
 59378.597 &        ---          &       ---     &  9.44\plm0.62 & 6.92\plm0.39  & 3.56\plm0.27 & 2.71\plm0.20 & 2.77\plm0.21 & 2.69\plm0.15	 \\
 59396.359 & 2.60\plm0.78$^4$  & -0.51\plm0.22 &  9.89\plm0.46 & 7.73\plm0.36  & 3.53\plm0.19 & 2.57\plm0.17 & 2.62\plm0.17 & 2.66\plm0.15	 \\
 59427.389 & 2.12\plm0.95  & -0.97\plm0.03 & 10.16\plm0.47 & 7.31\plm0.34  & 4.02\plm0.18 & 2.74\plm0.15 & 2.57\plm0.14 & 2.69\plm0.12	 \\
 59522.399 & 1.61\plm0.40  & -0.38\plm0.22 &  9.71\plm0.36 & 7.25\plm0.33  & 3.80\plm0.14 & 2.54\plm0.11 & 2.43\plm0.13 & 2.26\plm0.10	 \\
 59527.740 & 2.59\plm0.62  & -0.61\plm0.18 &  9.53\plm0.44 & 7.18\plm0.33  & 3.80\plm0.17 & 2.52\plm0.14 & 2.41\plm0.13 & 2.22\plm0.10	 \\
 59552.469 & 2.75\plm0.74  & +0.18\plm0.18 &  8.85\plm0.41 & 7.18\plm0.33  & 3.60\plm0.13 & 2.79\plm0.13 & 2.69\plm0.15 & 2.34\plm0.11	 \\
 59556.222 &        ---          &       ---     &	   ---     &	  ---	   & 3.84\plm0.36 & 2.76\plm0.18 &      ---	&      --- 	 \\
 59560.128 &        ---          &       ---     &  9.02\plm0.50 & 6.86\plm0.32  & 3.80\plm0.21 & 2.34\plm0.15 & 2.35\plm0.20 & 2.10\plm0.13	 \\
 59583.910 & 1.41\plm0.41  & -0.40\plm0.22 &  9.62\plm0.35 & 7.31\plm0.27  & 3.73\plm0.17 & 2.71\plm0.12 & 2.62\plm0.12 & 2.47\plm0.11	 \\
 59674.611 & 3.67\plm2.33$^4$  & -0.65\plm0.35 & 10.74\plm0.60 & 7.25\plm0.33  & 3.28\plm0.18 & 2.59\plm0.17 & 2.55\plm0.16 & 2.39\plm0.13   \\		
 59703.993 & 2.52\plm0.80$^4$  & -0.20\plm0.22 &  9.02\plm0.42 & 7.18\plm0.33  & 3.91\plm0.18 & 2.76\plm0.15 & 2.65\plm0.14 & 2.50\plm0.11   \\		
 59734.146 & 2.75\plm0.99  & -0.25\plm0.23 & 10.16\plm0.37 & 7.38\plm0.27  & 3.80\plm0.14 & 2.84\plm0.16 & 2.57\plm0.14 & 2.59\plm0.12   \\		       
 59764.390 & 5.00\plm0.53  & -0.54\plm0.12 &  9.98\plm0.37 & 7.38\plm0.27  & 3.91\plm0.14 & 2.71\plm0.12 & 2.53\plm0.11 & 2.47\plm0.09	 \\
 59795.644 & 2.96\plm0.70  & -0.38\plm0.21 &  9.98\plm0.37 & 7.31\plm0.27  & 3.80\plm0.14 & 2.66\plm0.12 & 2.39\plm0.13 & 2.34\plm0.11	 \\
 59883.310 & 4.65\plm0.44  & -0.63\plm0.18 &  9.52\plm0.45 & 7.45\plm0.28 & 3.67\plm0.17 & 2.37\plm0.13 & 2.53\plm0.17 & 2.24\plm0.11 \\
 59914.585 & 1.41\plm0.11$^4$  &  ---                 & 8.61\plm0.74 & 7.59\plm0.36 &  3.57\plm0.17 & 2.69\plm0.15 & 2.12\plm0.25 & 2.20\plm0.17 \\ 
 59919.955 & 6.27\plm3.41$^4$  &  ---                 &   ---              & 6.93\plm0.39 &  3.77\plm0.25 & 2.14\plm0.18 &  ---                & 2.48\plm0.21 \\
 59923.339 & 12.93\plm4.50$^4$ &-0.59\plm0.25  & 9.27\plm0.80 & 7.32\plm0.48 &  3.57\plm0.31 & 1.99\plm0.23 & 2.68\plm0.42 & 2.08\plm0.18 \\
 59936.658 & 4.49\plm1.21$^4$  & -0.79\plm0.20  & 9.98\plm0.47 & 7.59\plm0.28 &  3.99\plm0.19 & 2.62\plm0.15 & 2.46\plm0.16 & 2.50\plm0.14 \\
 59944.535 & 8.89\plm2.05$^4$  & -0.76\plm0.12  & 10.16\plm0.48& 7.39\plm0.28 &  3.60\plm0.17 & 2.82\plm0.16 & 2.55\plm0.17 & 2.46\plm0.14 \\
 59948.812 & 5.82\plm2.20$^4$  & -0.79\plm0.21  & ---                 & 6.99\plm0.26 &  3.57\plm0.17 & 2.43\plm0.16 &  ---                & 2.16\plm0.19 \\
 59949.630 & 9.92\plm1.39  & -0.83\plm0.16  & 9.18\plm0.61 &  7.32\plm0.34 &  4.17\plm0.24 & 2.62\plm0.17 & 2.51\plm0.22 & 2.39\plm0.16 \\
 59950.838 & 6.52\plm1.75$^4$  & -0.81\plm0.19  & 9.18\plm0.61 & 7.25\plm0.34  &  3.28\plm0.22 & 2.74\plm0.18 & 2.37\plm0.20 & 2.18\plm0.14 \\
 59951.623 & 8.64\plm2.03$^4$  & -0.70\plm0.18  & 8.77\plm0.58 & 6.67\plm0.38 &   3.95\plm0.22 & 2.43\plm0.19 & 2.46\plm0.21 & 2.26\plm0.15 \\
 59952.411 & 7.13\plm2.15$^4$  & -0.74\plm0.21  & 9.89\plm0.65 & 6.86\plm0.39 &  3.64\plm0.24 & 3.01\plm0.23  & 2.70\plm0.23 & 2.18\plm0.17 \\
 59953.011 & 6.43\plm2.03$^4$  & -0.64\plm0.21  & 8.22\plm0.54 & 7.66\plm0.36 &  3.32\plm0.22 & 2.20\plm0.17  & 1.96\plm0.21 & 1.93\plm0.15 \\
 59954.007 & 4.43\plm2.21$^4$  & -0.68\plm0.32  & 9.35\plm0.62 & 7.52\plm0.36 &  3.70\plm0.26 & 2.57\plm0.20  & 2.16\plm0.48 &  2.24\plm0.13 \\
 59955.596 & 6.49\plm2.15$^4$  & -0.51\plm0.22  & 10.64\plm0.60& 6.86\plm0.39 &  3.44\plm0.23 & 2.67\plm0.20 & 2.27\plm0.22 &  2.24\plm0.15 \\
 59956.868 & 8.37\plm2.52  & -0.40\plm0.21  & 9.98\plm0.66 & 7.59\plm0.36 &  3.77\plm0.25 & 2.67\plm0.20 &  2.05\plm0.20.&  2.20\plm0.15 \\
 59957.791 & 5.25\plm3.70$^4$  &  ---                 & 9.02\plm0.60 & 7.05\plm0.40 &  3.50\plm0.23 &  2.72\plm0.21 &  2.49\plm0.21 & 2.30\plm0.15 \\
 59964.619 & 6.62\plm1.70  & -0.26\plm0.23  & 9.02\plm0.51 & 7.12\plm0.33 &  3.95\plm0.22 & 2.62\plm0.20 &  2.25\plm0.19 & 2.35\plm0.13 \\
 59974.341 & 6.61\plm2.79$^4$  & -0.26\plm0.35  & 10.16\plm0.67& 6.86\plm0.39 &  3.81\plm0.25 & 2.55\plm0.19 &  ---                & 2.37\plm0.16 \\
 59975.276 & 5.03\plm1.67$^4$  & -0.33\plm0.25  & 9.53\plm0.72 & 7.32\plm0.27 &  3.84\plm0.18 & 2.57\plm0.15  &  2.53\plm0.22 & 2.48\plm0.14 \\
 59977.793 & ---                 &   ---                 & 11.25\plm0.75& 6.93\plm0.39$^4$ &  3.74\plm0.28 & 2.84\plm0.24 &  2.44\plm0.33 & 2.48\plm0.16 \\
 59984.472 & 4.67\plm2.09$^4$  & +0.16\plm0.37  & 9.27\plm0.71 & 7.81\plm0.44 &  3.77\plm0.29 & 2.69\plm0.23 &  2.60\plm0.22 & 2.50\plm0.17 \\
 59994.269 & 7.28\plm3.70$^4$  & ---                   & 10.07\plm0.67 & 7.88\plm0.37 &  3.60\plm0.24 & 2.87\plm0.22 &  2.37\plm0.23 & 2.46\plm0.16 \\
 60004.532 & ---                 & ---                   & 10.74\plm0.71 & 7.52\plm0.43 &  4.17\plm0.28 & 2.52\plm0.22 &  2.99\plm0.35 & 2.24\plm0.15 \\
 60009.765 & ---                 & ---                   & 	   ---             & 6.93\plm0.26 &  3.64\plm0.17 & 2.32\plm0.15 &  ---                 & 2.10\plm0.12 \\
 60014.532 & 2.79\plm1.15$^4$  & +0.09\plm0.31  & 9.18\plm0.52  & 7.25\plm0.34 &  3.84\plm0.22 & 2.59\plm0.17 &  2.18\plm0.19 & 2.46\plm0.14 \\
 60024.925 & ---                 & ---                    & 8.38\plm0.55 & 7.25\plm0.34 &  3.54\plm0.23 & 2.55\plm0.19 &  1.94\plm0.17 & 2.18\plm0.14 \\
 60034.588 & 3.52\plm1.15$^4$  & +0.73\plm0.25  & 9.10\plm0.43 & 7.12\plm0.27 &  3.57\plm0.17 & 2.59\plm0.17 &  2.48\plm0.21 & 2.06\plm0.12 \\
 60044.968 & ---                 & ---                   &  9.98\plm0.56 & 7.45\plm0.35 & 4.10\plm0.23 & 2.67\plm0.18 &  2.21\plm0.17 & 2.22\plm0.15 \\
 60054.563 & 2.78\plm1.44$^4$  & +0.06\plm0.42  & 9.89\plm0.65 & 6.86\plm0.39 &  3.08\plm0.20 & 2.64\plm0.23 &  2.56\plm0.22 & 2.22\plm0.15 \\
 60065.357 & 1.38\plm0.77$^4$  & +0.33\plm0.36  & 9.10\plm0.43 & 7.32\plm0.27 &  3.88\plm0.18 & 2.30\plm0.15 &  2.42\plm0.16 & 2.37\plm0.14 \\
 60095.205 & 4.22\plm1.60$^4$  & -0.02\plm0.31   & 8.85\plm0.42 & 8.03\plm0.45 & 3.91\plm0.26 & ---                 &  2.18\plm0.15 & --- \\
 60126.288 & 3.95\plm1.11$^4$  & -0.31\plm0.26   & 9.80\plm0.46 & 7.32\plm0.27 & 3.67\plm0.17 & 2.62\plm0.15 & 2.32\plm0.15 & 2.30\plm0.13 \\
 60156.127 & 1.89\plm0.99$^4$  & -0.46\plm0.54   & 9.36\plm0.53 & 6.55\plm0.31 & 3.70\plm0.21 & 2.67\plm0.18 & 2.42\plm0.16 & 2.30 \plm0.13 \\
\hline
\hline
\end{tabular}

$^{1}$ The observed 0.3-10 keV fluxes are given in units of $10^{-16}$ W m$^{-2}$. \\

$^{2}$ The hardness ratio is defined as $HR = \frac{hard - soft}{hard+soft}$ with the soft and hard bands in the 0.3-1.0 and 1.0-10 keV bands, respectively, appying the BEHR program by \citet{park2006}

$^{3}$ The Fluxes in the UVOT filters are given in units of $10^{-15}$ W m$^{-2}$. \\

$^4$ 0.3-10 keV fluxes estimated from the course rates based on the latest possible spectral analysis.

\end{table*}

\begin{table*}
\caption{Merged UVOT data as shown in Figure\, \ref{swift_uvot_lc}.  \label{uvot_merge} }
 \centering
  \begin{tabular}{ccrrrrrrr}
  \hline
  \hline

MJD Range &  MJD$_{\rm center}$ &  bin width$^1$ 
& $F_{\rm V }^{2}$ 
& $F_{\rm B }^{2}$ 
& $F_{\rm U }^{2}$ 
& $F_{\rm W1 }^{2}$ 
& $F_{\rm M2 }^{2}$ 
& $F_{\rm W2 }^{2}$  \\
\hline  
56595 - 57590 & 57092.5 & 497.5 &   9.22\plm0.26 & 7.23\plm0.21  & 3.56\plm0.10   &  2.59\plm0.08  &  2.53\plm0.08 & 2.37\plm0.07  \\
57762 - 57969 & 57865.5 & 103.5 &   9.24\plm0.26 & 7.07\plm0.21  & 3.67\plm0.11   &  2.76\plm0.10  & 2.56\plm0.10  & 2.39\plm0.07 \\ 
58065 - 58339 & 58202.0 & 137.0 &   9.30\plm0.18 & 7.07\plm0.21  & 3.64\plm0.11   &  2.63\plm0.08  &  2.58\plm0.10 & 2.43\plm0.07  \\
58422 - 58704 & 58563.0 & 141.0 &   9.40\plm0.18 & 7.20\plm0.21  & 3.74\plm0.11   &  2.70\plm0.08  & 2.54\plm0.07  & 2.43\plm0.07 \\ 
58788 - 59062 & 58925.0 & 137.0 &   9.57\plm0.27 & 7.25\plm0.21  & 3.92\plm0.11   &  2.73\plm0.08  & 2.61\plm0.10  & 2.43\plm0.10 \\
59154 - 59427 & 59290.5 & 136.5 &   9.46\plm0.27 & 7.25\plm0.21  & 3.74\plm0.11   &  2.68\plm0.11  & 2.54\plm0.10  & 2.47\plm0.10 \\
59527 - 59795 & 59661.0 & 134.0 &   9.51\plm0.27 &7.20\plm0.21   & 3.74\plm0.11   &  2.68\plm0.08  & 2.56\plm0.10  & 2.35\plm0.07 \\
59883 - 60156 & 60019.5 & 136.5 &   9.51\plm0.28 &7.29\plm0.21   & 3.71\plm0.11   &  2.55\plm0.10  & 2.45\plm0.10  & 2.20\plm0.09 \\
\hline
56595 - 60156 & 58375.5 & 1780.5  &  9.45\plm0.23 & 7.15\plm0.14  & 3.68\plm0.11   &  2.66\plm0.09  & 2.55\plm0.06  & 2.42\plm0.06 \\
\hline
\hline
\end{tabular}

$^{1}$ given in days \\

$^{2}$ The reddening-corrected UVOT fluxes are given in units of $10^{-15}$ W m$^{-2}$. \\

\end{table*}

\label{last page}

\end{document}